\documentclass[preprint,5p,times,twocolumn]{elsarticle}
\biboptions{sort&compress}
\usepackage[hidelinks]{hyperref}
\usepackage{lineno}
\usepackage{graphicx, amsmath, amssymb, xcolor, subcaption, enumitem}
\usepackage[rightcaption]{sidecap}

\renewcommand{\d}{\mathrm{d}}

\journal{Fusion Engineering and Design}

\begin{document}

\begin{frontmatter}

\title{Deuterium--Tritium Levitated Dipole Fusion Power Plants}
\author[openstar]{T. Simpson\corref{cor1}}\ead{tom.simpson@openstar.nz} 
\author[openstar]{R.A. Badcock}
\author[openstar]{T. Berry}
\author[openstar]{C.S. Chisholm}
\author[openstar]{P.J. Fimognari}
\author[openstar]{P. Fisher} 
\author[openstar]{D.T. Garnier} 
\author[openstar]{K. Lenagh-Glue} 
\author[openstar]{B. Leuw} 
\author[openstar]{R. Mataira}
\author[openstar]{L. Meadows}
\author[openstar]{T. McIntosh} 
\author[openstar]{J. Poata}
\author[openstar]{K. Richardson}
\author[openstar]{B. Smith}
\author[openstar]{A. Simpson}
\author[openstar]{J.D. Tyler}
\author[openstar]{T. Wordsworth} 

\affiliation[openstar]{organization={OpenStar Technologies Limited},
                       addressline={20 Glover Street, Ngauranga},
                       city={Wellington},
                       postcode={6035},
                       country={New Zealand}}
                       
\cortext[cor1]{Corresponding author}

\begin{abstract}
    Levitated dipole reactors offer an attractive path towards economic fusion power generation. The intrinsic decoupling of the confining magnetic field-generating REBCO magnets and the vacuum vessel offer unparalleled accessibility and maintainability, allowing for high plant duty factors and theoretically low electricity prices. In order to achieve rapid deployment of fusion power to the grid, the use of the Deuterium-Tritium (DT) fuel cycle is required due to its lower required plasma triple products. Historically, designs of levitated dipole fusion power plants have targeted advanced fuels as a DT device was seen to be infeasible due to the high fluxes of $14.1$~MeV neutrons on the superconducting core magnet. This study presents high level designs for two feasible first-of-a-kind (FOAK) DT levitated dipole fusion power plants, the larger of which produces $667$~MW of fusion power and is predicted to produce $208$~MW of net electric power. Both designs consist of a heavily neutron-shielded, high-field REBCO core magnet capable of producing peak magnetic field strengths of $23$~T while keeping peak mechanical strains below $0.4$~\%. The neutron shielding is comprised of a layered structure of tungsten and boron carbide, which allows for $92$\% of the heat deposited in the neutron shield to be radiated out to the first wall while still providing sufficient neutron attenuation to give adequate REBCO conductor lifetimes. The core magnet REBCO coil is comprised of a small ``sacrificial'' section and a larger semi-permanent section. The sacrificial section, comprising $\sim20\%$ of the coil, will have a neutron damage limited lifetime of $\sim1$~year, after which the core magnet will be quickly removed from the vacuum vessel and replaced. This allows the damaged core magnet to be refurbished and reused, reducing cost and allowing for economic fusion power generation from a DT levitated dipole reactor.
\end{abstract}


\end{frontmatter}



\section{Introduction}\label{sec:intro}

\begin{figure}[!b]
    \centering
    \includegraphics[width=\linewidth]{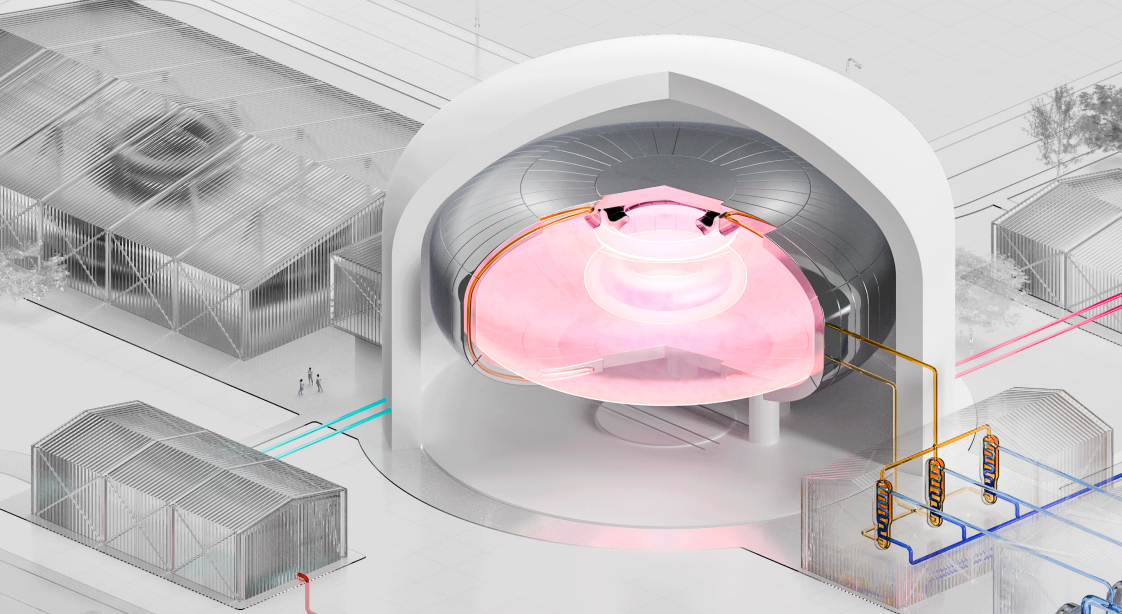}
    \caption{Artistic render of a first-of-a-kind levitated dipole reactor. The core magnet pictured within the plasma in the center of the image is levitated in a large, simple two layer vacuum vessel. The levitation force is provided by a smaller magnet mounted at the top of the inner vacuum vessel. The inner vacuum vessel is surrounded by a tritium breeding blanket. The outer vacuum vessel is constructed from reinforced concrete and is designed to handle the atmospheric forces required for such a large vacuum vessel. Supporting systems and infrastructure are also depicted.}
    \label{fig:reactor_schematic}
\end{figure}

\begin{figure*}[t!]
    \centering
    \includegraphics[width=1\linewidth]{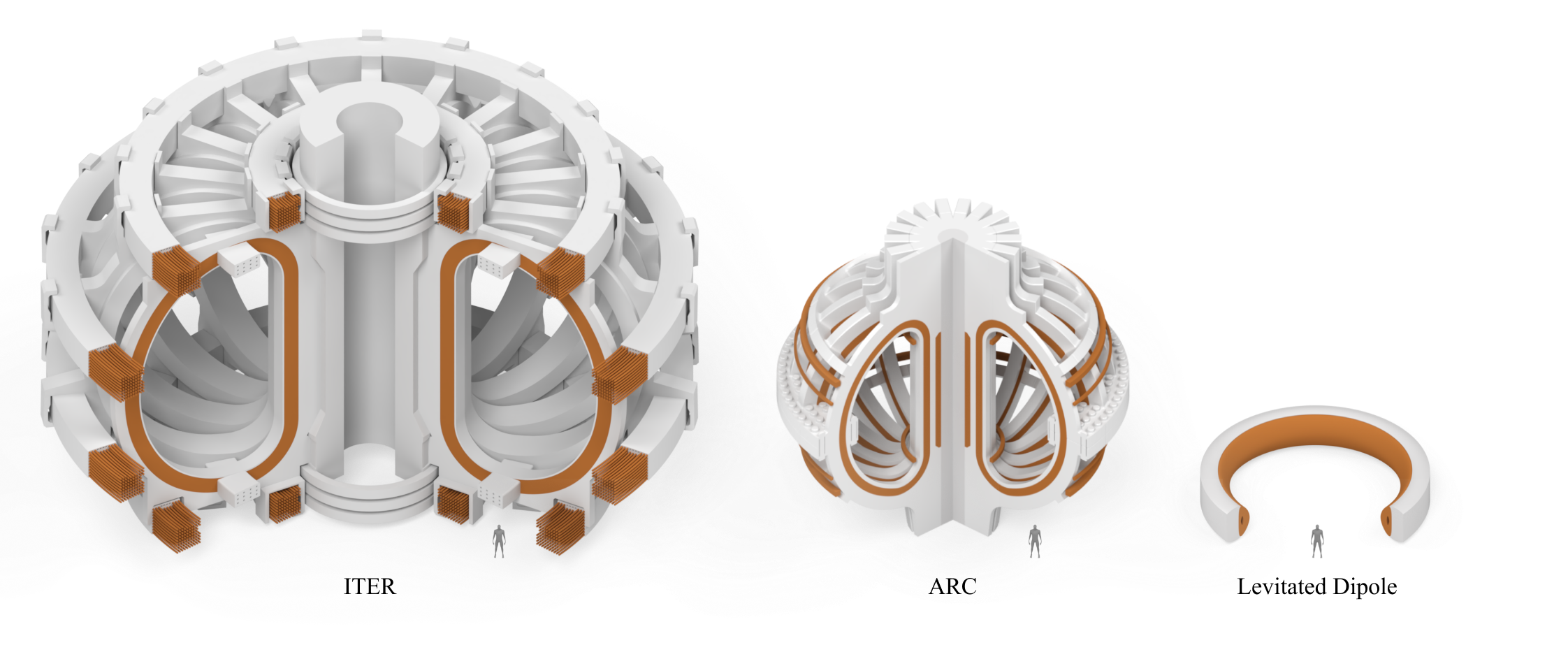}
    \caption{A comparison of scale between the core magnet of a $\sim 500$ MW thermal power levitated dipole and the magnet systems of similar power output tokamaks. The vacuum vessel of a levitated dipole is approximately twice the diameter of ITER's outer vacuum vessel, however, it has been excluded from this diagram as the magnet systems are the highest cost component of any magnetic confinement concept.}
    \label{fig:magnet_scale}
\end{figure*}

The need for a scalable base load power generation source has been identified as a key requirement for addressing climate change and the rapid growth of global energy demands \cite{IPCC:2023, wolfram:2012, mauleon:2022, vries:2023}. Fusion offers unparalleled scalability due to the energy density and abundance of Deuterium. However, to make a significant impact in the global energy market, the roll out of early fusion power plants will need to be rapid, which requires competitive energy prices. Recent studies into the economics of a fusion power plant \cite{maris:2024} show that plasma disruptions pose a major risk of raising the price of electricity, highlighting the benefit of disruption-free configurations. Additionally, the large external coils required for tokamaks and stellarators makes maintenance intrinsically difficult --- usually requiring the complete or partial disassembly of the reactor \cite{sorbom:2015, MANTA:2024, lion:2025} --- impacting plant availability and resulting in an increased price of electricity. Commercial fusion power plants will need to treat accessibility and maintainability as a priority.

First proposed by Akira Hasegawa in 1987 after early observations of planetary magnetospheres \cite{hasegawa:1987}, levitated dipoles benefit from favorable physics and engineering properties that merit their investigation as fusion power plants. Although subsequent levitated dipole experiments have replicated many of the attractive features of these plasmas in a laboratory setting \cite{boxer:2010, saitoh:2011, goto:2006}, little research has been carried out on their performance as fusion energy devices. Levitated dipoles are characterized by a single plasma confining superconducting coil, the `core magnet', levitated in the center of a large vacuum vessel as shown in Fig.~\ref{fig:reactor_schematic}. The levitation force and position control is provided by relatively weak poloidal field coils mounted outside of the inner vacuum vessel. In the simplest case, which we assume in this study, there will only be one external poloidal field coil which we name the `top magnet'. This configuration of magnets and vacuum vessel does not require any complex interlocking of components, allowing for a level of access and maintainability unique among magnetically confined fusion devices. Hence, the maintenance, replacement, and iteration of key components, such as the core magnet, vacuum vessel, and tritium breeding blanket can be fast and completed with simpler robotic systems than possible in other fusion concepts. 

Operating a superconducting coil in the core plasma region introduces engineering challenges that must be accounted for in a commercial device. Previous studies have attempted to mitigate these challenges by focusing primarily on reactors using advanced fuel cycles \cite{hasegawa:1990, kesner:2003}. However, the order of magnitude increase in the required triple product to reach ignition compared to the deuterium--tritium (DT) fuel cycle places demanding requirements on the plasma performance, increasing the size of the device beyond what would be acceptable for a first-of-a-kind (FOAK) fusion power plant. Therefore, this study focuses on the design of levitated dipole reactors using the DT fuel cycle and tackles the key challenges introduced by the high flux of $14.1$ MeV neutrons. In this study we show that the use of a DT fuel cycle allows for levitated dipole reactors with smaller magnetic systems than comparable fusion power output tokamaks, as shown in Fig.~\ref{fig:magnet_scale}. This in turn can be leveraged along with the inherent accessibility to allow for appealing power plant economics.

The structure of this study aims to offer a concise introduction to the aspects of a levitated dipole reactor that differ from other magnetically confined fusion concepts. Section~\ref{sec:design_principles} discusses the equilibrium and stability of the dipole plasma and details the engineering of a levitated dipole reactor and the considerations necessary to satisfy the requirements of a viable FOAK fusion power plant. Section~\ref{sec:methods} describes the workflow and optimizer used to find viable operating points. This work presents for the first time a dipole magnet design that has been optimized for fusion performance within demonstrated engineering limits \cite{hartwig:2024}.  Section \ref{sec:analysis} then presents two design points and provides an analysis of key parameters.

\begin{SCfigure*}[0.67][t]
    \centering
    \includegraphics[width=1.5\linewidth]{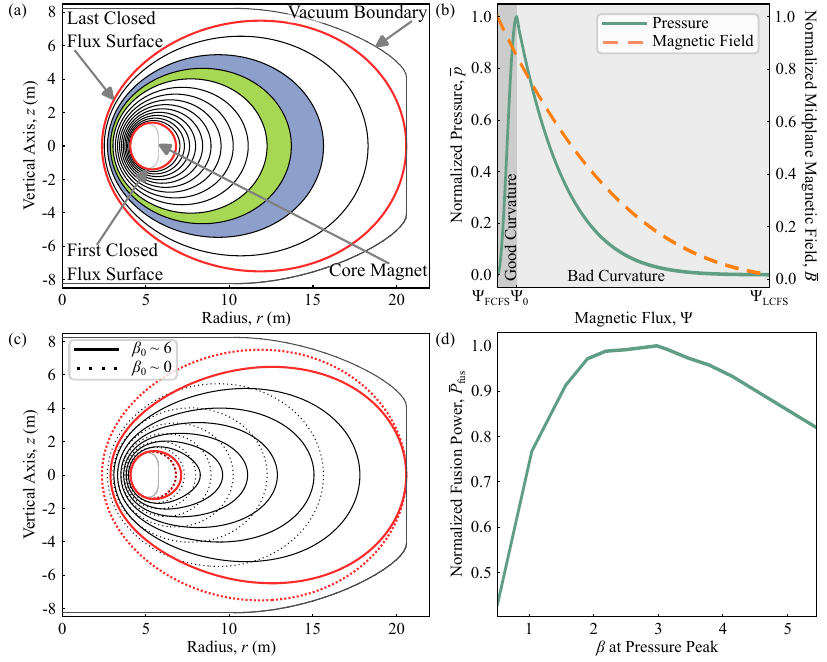}
    \caption{(a) Radial cross section showing the poloidal magnetic flux of a fusion power plant scale dipole at low $\beta$. The first closed flux surface is determined by the core magnet cryostat shell which acts as an inner limiter and the last closed flux surface is set by the vacuum vessel which acts as an outer limiter. The blue and green flux-tubes contain equal amounts of poloidal magnetic flux but the blue flux tube encloses a larger volume. When the two flux tubes interchange position, the volume of the blue flux-tube decreases and the plasma within is compressed and heated in accordance with Eq.~\eqref{eq:pvgamma}. Simultaneously, the volume of the green flux-tube increases and the plasma within expands and cools. (b) Pressure (green solid line, left axis) and magnetic field at the midplane (dashed orange line, right axis) from the equilibrium shown in (a) as functions of poloidal magnetic flux, $\Psi$. (c) Radial cross section showing the poloidal magnetic flux at high $\beta$ (solid lines) overlayed with low $\beta$ contours from (a) (dotted lines). (d) Total fusion power as a function of $\beta_0$ with the pressure peak location adjusted to ensure it remains two $\alpha$ gyro-orbits away from $\Psi_{\rm fcfs}$. The plasma expansion observed in (c) results in the maximum fusion power being achieved at $\beta_0\sim3$.}
    \label{fig:eqbs}    
\end{SCfigure*}


\section{Design Principles}\label{sec:design_principles}

\subsection{Dipole Physics Basis}\label{sec:dp_plasma}

The levitated dipole mimics the plasma confinement of planetary magnetospheres, where a dipole magnetic field creates the environment for stable, steady state, and centrally peaked plasmas \cite{schulz:1974, lyon:2000, birmingham:1969, falthammar:1965}. The key difference in a levitated dipole is, as the name suggests, the levitation of the core magnet. Removing the physical supports creates a region of closed flux surfaces and eliminates plasma losses along field-lines. Therefore, the plasma in a levitated dipole forms in this region of closed flux surfaces between the core magnet and the limiting vacuum vessel, and its equilibrium is governed by the reduced (toroidal-field-free) Grad-Shafranov equation \cite{garnier:1999}. Therefore unlike most magnetically confined plasmas there is both a last closed flux surface, $\Psi_{\rm lcfs}$, created by either a limiter on the vacuum vessel or a separatrix, and a first closed flux surface, $\Psi_{\rm fcfs}$, caused by the plasma limiting on the core magnet. Between these surfaces lies a single pressure peak located at the flux surface denoted by $\Psi_0$. Another key difference is that particles orbiting interior to $\Psi_0$ travel in a region of absolute good curvature, denoted by $\mathfrak{R}_{\rm GC}$, and those orbiting beyond do so in a region of absolute bad curvature, denoted by $\mathfrak{R}_{\rm BC}$. The plasma in a levitated dipole can either be diverted or limited anywhere around $\Psi_{\rm lcfs}$, with the outer midplane offering the lowest wall loadings. In this work, we have elected to study a dipole configuration that is limited on the outer midplane. This  merely serves as a test of feasibility and future designs will focus on diverted dipole plasmas.

Fundamentally, the core magnet and some form of plasma heating are all that is required to create a fusion plasma with a levitated dipole. Therefore, any particular levitated dipole design can be primarily characterized by four parameters: the major radius of the core magnet coil, $R_{\rm m}$;  the core magnet coil aspect ratio, $R_{\rm m} / a_{\rm m}$, with $a_{\rm m}$ the core magnet minor radius; the total core magnet current, $I_{\rm tot}$; and the effective plasma aspect ratio, $\delta V_{\rm lcfs} / \delta V_{\rm 0}$, defined using the ratio of the differential flux tube volumes at $\Psi_{\rm lcfs}$ and $\Psi_0$, respectively. As we show in Section~\ref{sec:dp_stability}, the plasma aspect ratio is the key factor that drives the peaked pressure profiles required for fusion.

\subsubsection{Equilibrium and Stability}\label{sec:dp_stability}

Theory has predicted \cite{rosenbluth:1957, hasegawa:1987} and observations have shown \cite{boxer:2010, yoshida:2010} that in $\mathfrak{R}_{\rm BC}$ a levitated dipole plasma equilibrium, an example of which is given in Fig~\ref{fig:eqbs}, is achieved at the limit of marginal stability to interchange modes:
\begin{equation}
    \delta (p\delta V^\gamma) = 0,
    \label{eq:pvgamma}
\end{equation}
where $\delta V=\oint\d l/ B$ and $p$ are the differential flux tube volume and plasma pressure, respectively. This results in a critical pressure gradient:
\begin{equation}
    d \equiv -\frac{\d \ln p}{\d \ln \delta V} < \gamma,
    \label{eq:d}
\end{equation}
which defines the MHD stability limit and, when exceeded, results in the formation of large scale convective cells that act to transport energy out to the last closed flux surface \cite{kesner:2000-2, rey:2001, hassam:1979, shukla:1984}. Convective cells can also form in a marginally stable profile which act to transport particles to and from the core without the net transport of energy \cite{pastukhov:1992}. Therefore, in a levitated dipole where $B\propto1/r^3$ this results in the following scaling of the peak plasma pressure:
\begin{equation}
    p_0 = p_{\rm lcfs}\left(\frac{\delta V_{\rm lcfs}}{\delta V_{\rm 0}}\right)^{\gamma} \propto \left(\frac{R_{\rm lcfs}}{R_0}\right)^{20/3},
    \label{eq:p_ratio}
\end{equation}
which heavily incentivizes the use of a relatively small magnet in a large vacuum chamber. A similar critical gradient for the temperature and density can also be obtained:
\begin{equation}
    \eta \equiv \frac{\d \ln T}{\d \ln n} = \gamma-1,
    \label{eq:eta}
\end{equation}
which describes the stability to drift frequency ``entropy'' modes \cite{kesner:2000, kesner:2002, simakov:2001} and is valid for arbitrary plasma $\langle \beta \rangle$ \cite{simakov:2002}. It has been shown in pellet injection experiments \cite{garnier:2017} that changing the density gradient will lead to either an outward particle flux when $\eta < 2/3$ or an inward particle flux when $\eta > 2/3$ which act to return the system to the $\eta=\gamma-1$ profile.


\subsubsection{Pressure Peak Location}\label{sec:dp_peak_location}

In steady state the pressure peak, $\Psi_0$, is the point that allows for power balance to be achieved in both the good and bad curvature regions of the plasma. The transport in $\mathfrak{R}_{\rm BC}$ has been characterized to some extent \cite{boxer:2010, garnier:2017}, however, the transport at the plasma edges and in $\mathfrak{R}_{\rm GC}$ is unknown. The lack of degrading modes in $\mathfrak{R}_{\rm GC}$ suggests that the transport in this region could approach classical, allowing for very steep temperature and density gradients. However, the prompt $\alpha$ particle losses will eventually dominate at a distance proportional to the Larmor radii of the energetic $\alpha$ particles. The low magnetic field strengths at the outboard side of the core magnet result in prompt $\alpha$ particle losses becoming the practical limit on the location of $\Psi_0$. The plasma equilibria in this study have $\Psi_0$ located at the prompt $\alpha$ loss limit. Determining whether or not defining $\Psi_0$ in this way would result in a self-consistent equilibrium requires knowledge of the energy transport in $\mathfrak{R}_{\rm GC}$. This particular problem is out of the scope of this study and will be a focus of future levitated dipole experiments. Furthermore, we assume the heat conducted to $\Psi_{\rm fcfs}$ to be negligible ($P_{\rm \kappa, in}=0$) requiring all heat deposited in $\mathfrak{R}_{\rm GC}$ to be radiated away \cite{kesner:1997, pastukhov:1992}. 


\subsubsection{$\beta$ Limit}\label{sec:dp_large_beta}
Due to the extremely peaked nature of the pressure profile as shown in Fig.~\ref{fig:eqbs}, it is useful to quantify both the global $\beta$:
\begin{equation}
    \label{eq:beta_p}
    \langle \beta \rangle \equiv 2 \mu_0 \frac{\langle p \rangle}{\langle B^2 \rangle},
\end{equation}
where the average is taken over the plasma volume, and the local $\beta$ at the low field side pressure peak:
\begin{equation}
    \label{eq:beta_loc} 
    \beta_0 \equiv 2\mu_0\frac{p_0}{B_0^2}.
\end{equation}
$\beta_0$ has to be finite in order to confine the plasma, which for a particular equilibrium solution will translate to an equivalent limit on $\langle \beta \rangle$. A more practical limit on $\langle \beta \rangle$ is a result of the plasma expansion that follows increasing $p_0$, which results in an outwards movement of the poloidal flux lines. This effect can be seen by comparing the low $\beta_0$ and high $\beta_0$ equilibria in Fig.~\ref{fig:eqbs}(c). Fig.~\ref{fig:eqbs}(d) then shows the evolution of the fusion power as a function of $\beta_0$. As $\beta_0$ is increased the plasma pressure increases initially, however beyond a certain point the expansion results in a reduction of the peak pressure at the core according to Eq.~\eqref{eq:p_ratio}. This peak typically happens around $\beta_0$ values of $1-2$.  However, the total stored energy in the plasma is still increasing at these values of $\beta_0$, which causes the fusion power to peak at a higher value of $\beta_0$, in the range of $\sim3$ for many reactor configurations. 
    
\subsubsection{Plasma Edge Conditions}\label{sec:dp_edge}

Following on from Eq.~\eqref{eq:pvgamma}, the pressure in the plasma core is determined by both the device aspect ratio, $R_{\rm lcfs} / R_0$, as in Eq.~\eqref{eq:p_ratio} and the pressure at $\Psi_{\rm lcfs}$, denoted as $p_{\rm lcfs}$. For a given core pressure, $p_0$, higher values of $p_{\rm lcfs}$ will allow for smaller vacuum vessels in proportion with Eq.~\eqref{eq:p_ratio}, reducing the overall cost of the plant. Hence there is an incentive to design reactors with the highest possible edge pressures.

The physics defining an upper bound on the value of $p_{\rm lcfs}$ is not well understood as no dipole experiments have yet had enough heating power to generate edge conditions applicable to fusion power plants. However, gyrokinetic simulations have shown promising results for zonal flow formation in certain collisionality regimes \cite{ricci:2006a, kobayashi:2009, hoffmann:2023}, suggesting the possibility of edge or internal transport barrier formation. It is also believed that ambipolarity due to preferential scrape-off of the large gyro radius ions near the plasma edge could result in shear flows which in turn could lead to edge pedestal formation. As such, for the purpose of this study we have assumed that pedestal-like edge conditions are possible \cite{terry:2007, whyte:2010} and that plasma values at the plasma edge, denoted as $x_{\rm lcfs}$, are taken to be the values at the pedestal. The presence of an edge pedestal will be confirmed with results from a fusion relevant levitated dipole device. 

For this study we are considering an equilibrium limited on the outboard midplane. We do not expect this configuration to be able to create the edge conditions discussed in this section, however, this should not affect the engineering of the remaining reactor components. We anticipate that the higher performance assumed above will be achievable by adding shaping coils and diverting the equilibrium, the details of which are left for a future study.

\subsubsection{Energy Confinement Time}\label{sec:dp_intro_scaling}

The goal of this study is to show the viability of a levitated dipole fusion power plant. Part of this analysis involves modeling the required auxiliary heating power and its effect on the overall power balance. This in turn requires the notion of an energy confinement time which is typically extrapolated from experimental data \cite{luce:2008}. In this case, attempting to scale the energy confinement time from existing levitated dipoles would be counterproductive as the large difference in device scale required for a reactor will lead to unacceptable errors in the predicted performance. In lieu of experimental data, this study takes the reverse approach of assuming a confinement time for a reactor and then using Bohm-like and gyro-Bohm-like scaling to generate performance requirements for a small scale demonstration device. 

However, it is worth noting the differences in the construction of the confinement time in relation to those used by tokamaks and stellarators. The conductive losses in a levitated dipole are characterized by losses inwards towards the first closed flux surface, $P_{\rm \kappa, in}$, and losses out towards the last closed flux surface, $P_{\rm \kappa, out}$. In a levitated dipole, both mechanisms would contribute their own energy confinement time which could then be combined to reconstruct a global energy confinement time. However, due to the good curvature confinement in $\mathfrak{R}_{\rm GC}$ we expect the transport to approach classical limits. Therefore, we also expect losses out to the last closed flux surface to dominate ($P_{\rm \kappa, out} >> P_{\rm \kappa, in}$), implying the definition of global energy confinement time is generally well approximated by transport only in $\mathfrak{R}_{\rm BC}$. Furthermore, as outlined in Section~\ref{sec:dp_peak_location} we have assumed for simplicity that all energy deposited in $\mathfrak{R}_{\rm GC}$ is balanced by radiation alone ($P_{\rm \kappa, in} =0$). For this reason, we define the energy confinement time using only losses through $\Psi_{\rm lcfs}$ for the remainder of this study.

\subsection{Engineering Considerations}\label{sec:dp_eng}

One key advantage of the levitated dipole concept is the relative simplicity of the confinement magnet geometry, consisting of only poloidal field coils, in contrast to other magnetic confinement fusion systems. This simplicity would allow a dipole fusion power plant to be inherently maintainable and simple to manufacture at scale, enabling the rapid roll-out and iteration needed to make a significant impact in the electricity market. However, the cost of this overall system simplicity is the extreme operating environment experienced by the levitating high field core magnet. The core magnet must function without any physical connection to external systems for extended periods of time while  being surrounded by the fusing core plasma. The engineering and radiation shielding of the core magnet is the main focus of this study. The remaining systems are not spatially constrained nor coupled to the core magnet, resulting in greatly reduced complexity and risk.

\begin{figure}[t!]
    \centering
    \includegraphics[width=1\linewidth]{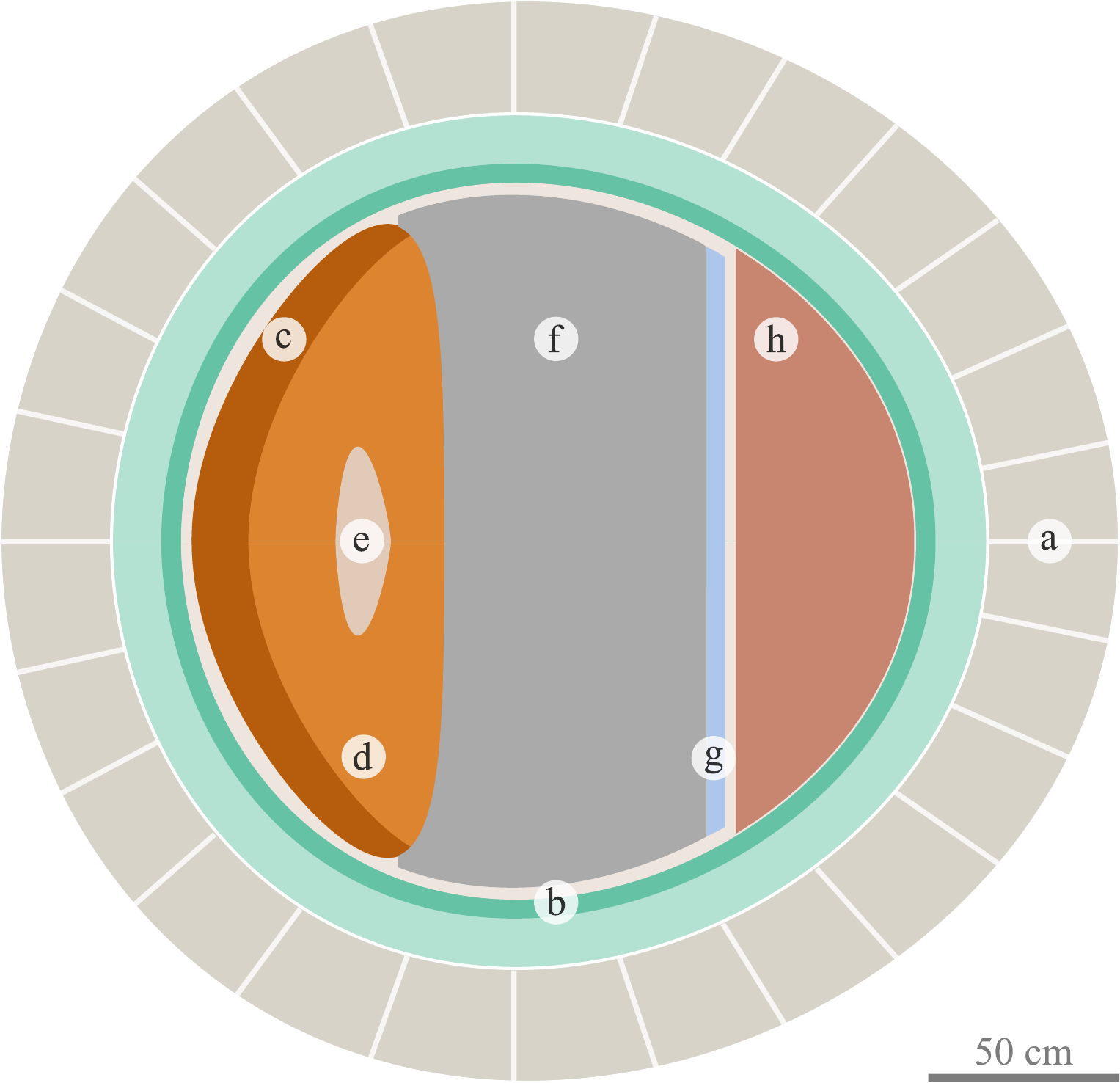}
    \caption{Representative cross section of a DT fusion power plant levitated dipole core magnet. (a) Radiatively cooled tungsten tiles. (b) ${\rm B}_4{\rm C}$ and WC neutron shield structure. (c) Sacrificial section of the REBCO coil where the neutron flux is higher. (d) Permanent section of the REBCO coil. (e) Low-field region. (f) Hoop stress retaining structural overband. (g) Cryogenic slush reservoir. (h) Reservoir for cooling the neutron shield structure.}
    \label{fig:magXsection}
\end{figure}

Fig.~\ref{fig:magXsection} depicts a representative cross section of a DT fusion-power-plant levitated-dipole core magnet. There are three main regions to the core magnet: the cryogenic REBCO coil and structural support (c-f), the coolant reservoirs (g-h), and the neutron shield (a-b). Due to the sensitivity to plasma radius in Eq.~\eqref{eq:p_ratio}, there are implied spatial constraints on each of these regions. The equilibrium physics outlined in Section~\ref{sec:dp_plasma} allow for steady-state reactor operation. However, as the core magnet is physically disconnected from all systems that would traditionally provide the required cooling power, the operation of a levitated-dipole fusion power plant must be pulsed to allow periodic removal of heat from the core magnet. This same disconnection of the magnet systems from the vacuum vessel enables simple maintenance and/or replacement of the core magnet and other key reactor components at the end of their operational lifetime. The aim of the designs presented in this study is to maximize the plant duty cycle through the following means: a latent-heat based cryogenic reservoir to minimize downtime, an on-board superconducting power supply with energy storage to remove the need to recharge the magnet when docked, and a high-performance neutron shield to minimize heating and damage from fusion neutrons.  

\subsubsection{HTS Coil Design}\label{sec:dp_hts_coil_design}

A simple design for the REBCO coil in a levitated dipole would have a rectangular \cite{kesner:2003, garnier:2006-2} or circular cross section. Although simple to manufacture, these geometries are not efficient in producing the required poloidal magnetic flux needed to effectively confine a dipole plasma. In the case of the rectangular cross-section coil, the corners of the coil will intersect with lines of poloidal flux, reducing the number of closed flux surfaces and increasing the flux tube volume, $\delta V_0$, at the pressure peak. In accordance with Eq.~\eqref{eq:p_ratio}, this then results in a lower peak pressure, $p_0$, which can only be compensated for by increasing the magnet strength. This is not the case for a circular cross section coil, instead the high curvature on the in-board side will result in high peak magnetic field strengths for a given amount of poloidal flux. In both cases, high peak magnetic field strengths are required which can result in untenable strains in the REBCO tape \cite{kesner:2003}. Furthermore, these designs would not allow for any magnetically sensitive systems, as described in Section~\ref{sec:dp_fp}, to be mounted on-board the core magnet. 

The design of the REBCO coil in this study alters the cross-sectional profile to minimize the induced mechanical stresses, through the reduction of peak magnetic field strengths, whilst simultaneously maximizing the plasma performance. However, the coil also needs to produce adequate magnetic field strengths in the region of highest $\beta_0$, located on the outboard side of the core magnet, in order to achieve higher plasma pressure as discussed in Section~\ref{sec:dp_large_beta}. Typically, magnets with a high aspect ratio ($R_{\rm m} / a_{\rm m}$) maximize this field strength. However, this comes at the expense of higher peak magnetic field strengths, and therefore stress, for a given total magnet current $I_{\rm tot}$. The coil cross section is therefore defined as the shape that offers the optimal tradeoff between these two competing goals. The method used to find this shape is described in Section~\ref{sec:m_hts_coil_optimisation}. 

In order to accommodate magnetically sensitive equipment on-board the core magnet, there must be a region of low magnetic field. Passive magnetic-shielding materials are only effective up to $\sim 100$ mT, making them not suitable in this case. The solution is to create a region within the REBCO coil whose shape is chosen such that the REBCO coil acts as its own magnetic shielding. This method has been demonstrated to work in OpenStar's Junior device \cite{chisholm:2026} which utilizes an arrangement of 14 REBCO coils to both generate the confining field and produce a low-field region to house the superconducting power supply. 

Unlike the Junior core magnet, the core magnet coil designs presented in this study utilize a cable-in-conduit conductor (CICC) architecture instead of simple non-insulated pancake style coils. The decision to move from pancake coils to a CICC architecture is motivated by the increased stiffness allowed by CICC, which in turn allows for Lorentz loads to be more easily transferred to an external structural over-band. The addition of an over-band will ultimately reduce the strain experienced by the REBCO tape, enabling high-field magnet designs. 

The strict spatial constraints make shielding the REBCO coil from the $14.1$~MeV DT neutrons a challenging proposition. To this end, the REBCO coil is split into two regions as shown in Fig.~\ref{fig:magXsection}: A small region of ``sacrificial'' REBCO conductor and the remaining semi-permanent section. The sacrificial section will experience higher neutron fluxes due to the thinner neutron shield and will therefore have a shorter operational lifespan, however it will also act as additional shielding for the remaining permanent REBCO conductor. The sacrificial portion will be demountable from the rest of the magnet so that it can be replaced at regular intervals without much difficulty. Having this region of sacrificial REBCO conductor in the coil reduces the performance requirements on the neutron shield allowing it to be thinner and hence increasing the magnetic field strength applied to the plasma and allowing higher plasma pressures.

\subsubsection{Superconducting Power Supply}\label{sec:dp_fp}

While required for the economic viability of a levitated dipole fusion power plant, the use of REBCO tape in the core magnet coil introduces the issue of resistive losses. There is no process for creating a superconducting joint with performance equal to that of a commercially available REBCO tapes~\cite{park:2014, kim:2013}. Therefore, the coil requires the use of normal-conducting jointing methods which will result in the dissipation of the current in the core magnet. This means that some form of power supply, including its auxiliary systems and energy source, are required to be installed on-board to compensate for the ohmic losses and enable the quasi-persistent operation of the core magnet coil. 

\begin{figure}[t!]
    \centering
    \includegraphics[width=1\linewidth]{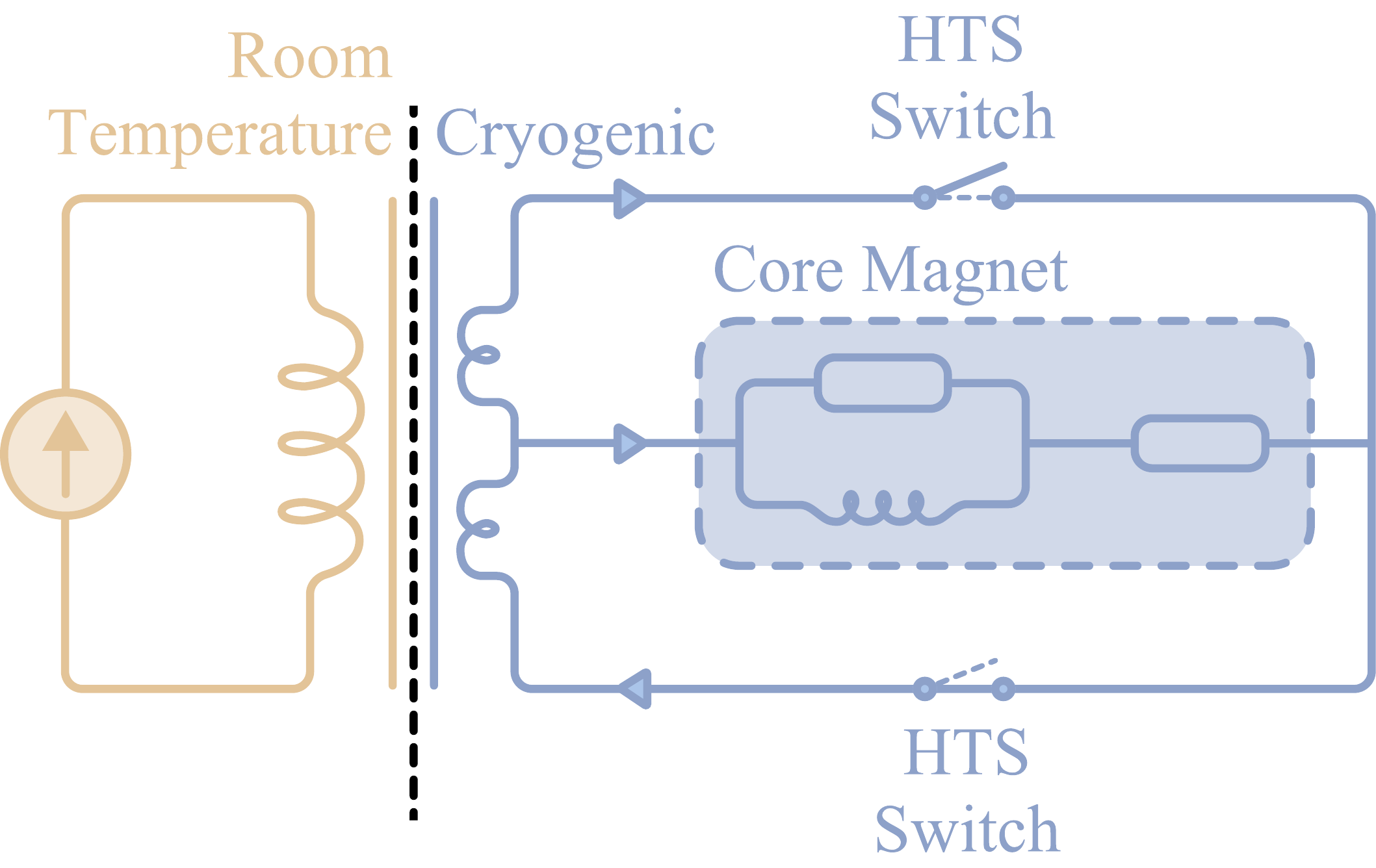}
    \caption{Circuit diagram of a center-tapped transformer-rectifier type superconducting power supply. On the primary side (orange) are the primary windings and the power electronics needed to drive the power supply. A transition from room temperature to cryogenic conditions occurs accross the transformer. The secondary side (blue) consists of a pair of secondary windings, a pair of superconducting switches, and the core magnet itself.}
    \label{fig:ctr_fp_circuit_diagram}
\end{figure}

Superconducting transformer rectifiers \cite{Leuw:2022,Geng:2025}, as shown in Fig.~\ref{fig:ctr_fp_circuit_diagram}, are a type of superconducting power supply, colloquially called ``flux pumps" \cite{VANDEKLUNDERT:1981,Wen:2022, Hoffman_2011}, which make use of both superconducting circuitry and switch elements \cite{gong:2025} to efficiently generate and maintain large currents in superconducting loads. Several circuit topologies exist, but they all make use of a step-down transformer to convert small amplitude AC currents from the primary side to large DC currents on the superconducting secondary side. Once the AC currents are in the superconducting circuit they can be rectified and driven into the core magnet coil over many cycles. There are a range of superconducting switching technologies, but currently the mechanism that provides the greatest switching action is a $J_{\rm c}(B)$ switch \cite{badcock:2022}. The energy source for the superconducting power supply will be either batteries or a bank of capacitors. Batteries offer higher energy densities than capacitors, however they could be more susceptible to radiation damage \cite{leita:2024}. In either case, both solutions will be recharged when the magnet is docked.

Superconducting power supplies are not high-power devices but rather excel at minimizing heat load in cryogenic high-current applications by transforming energy into the optimal current and voltage states \cite{Hamilton:2023}. This is important for the commercial viability of levitated dipoles as it means that there is an efficient method of compensating for the energy loss in the core magnet during operation. Without this mechanism, the core magnet current would decay during operation and require an additional period for re-charging during its down time, which will be a significant overhead on the operational duty cycle of the dipole.

\subsubsection{Cryogenic Cooling}\label{sec:dp_cryogen}

A cryogenic reservoir is required to keep the REBCO conductor in the core magnet coil in its superconducting state when operating in the levitated position. A traditional method to cool REBCO magnets is to run a cryogen loop and rely on the magnet's specific heat capacity and a constant flow of coolant to keep the magnet cold. In a levitated dipole, a constant flow is not possible due to the need to operate without physical connections to the rest of the plant, resulting in an operating mode where the dipole is allowed to warm up to a maximum temperature during operation before docking and re-cooling during a servicing period or down time. However, this approach limits the duty cycle of the plant to the speed at which the magnet can be re-cooled. For a large magnet with significant thermal mass, this is a lengthy process which will negatively affect the economic viability of the dipole by greatly reducing duty cycle. 

The cooling strategy proposed in this study is to use a cryogenic solid-liquid slush housed on-board the dipole that melts at a constant temperature. During operation, the latent heat of the slush provides the dipole with a constant-temperature thermal reservoir. This is doubly important for the dipole because it means the REBCO coil will not need to be designed with additional operational headroom to allow for a changing operating temperature. Once melted, the resulting liquid cryogen can be quickly pumped out of the reservoir and replaced with a new batch of slush cryogen. This can be achieved in a closed loop process such that the liquid cryogen can be re-processed into slush and stored in an external reservoir for future use.

There are two viable choices of cryogen for a high field REBCO magnet: Neon with a melting point of $24.6$~K \cite{ekin:2006}, and hydrogen with a melting point of $14.0$~K \cite{ekin:2006}. The reactors designed in this study use Neon as the cryogen due to its superior volumetric latent heat capacity, however, the lower temperatures and cost of hydrogen could make it an appealing choice for future reactors. Ultimately, the driving parameter for an economical dipole is to increase the operating duty cycle. The time it takes to replace a slush cryogen reservoir can be much reduced compared to re-cooling large magnet structures.

\subsubsection{Neutron Shielding}\label{sec:dp_Neutron_Shield}

\begin{figure}[t!]
    \centering
    \includegraphics[width=1\linewidth]{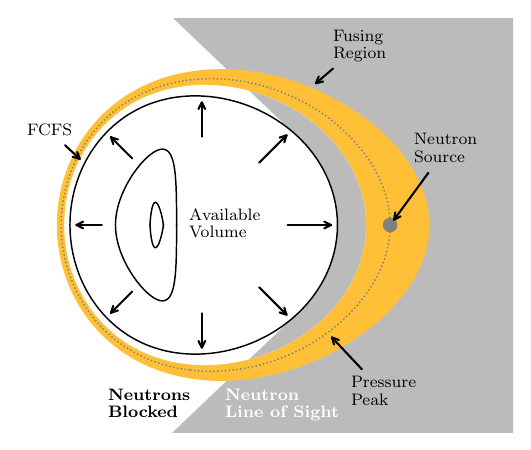}
    \caption{Sketch depicting the effective line of sight of neutrons emitted from the plasma. The core magnet takes up a small portion, resulting in only $\sim 25$~\% of neutrons passing through the $\Psi_{\rm fcfs}$ surface. The neutron shield is then defined to take all necessary space up to $\Psi_{\rm fcfs}$ to maximize performance.}
    \label{fig:neutron_fov}
\end{figure}

The problem of shielding the core magnet from fusion neutrons is aided by the geometry of the device. As shown in Fig.~\ref{fig:neutron_fov}, the core magnet only takes up a small portion of the field of view from the neutron source. The fraction of neutrons that pass through the space occupied by the core magnet was calculated to be $\sim25$~\% using OpenMC for a wide range of plasma equilibria. This incident fraction makes the problem of neutron flux attenuation comparable in difficulty to shielding a central column in a tokamak \cite{windsor:2021}, and hence similar shield materials are considered. The neutron shield thickness is optimized by defining the outer surface to coincide with the $\Psi_{\rm fcfs}$ contour. This gives ample room for shielding on the outboard side where flux expansion results in higher total fusion rates than in the core magnet bore (See Fig.~\ref{fig:neutron_flux} for more details). However, the only way to increase the neutron shield thickness in the core magnet bore is to move $\Psi_{\rm fcfs}$ further out from the core magnet coil. This then moves the pressure peak contour, $\Psi_0$, further out, incurring a steep fusion performance penalty according to Eq.~\eqref{eq:p_ratio}. Therefore, the choice of shielding material should prioritize efficient attenuation length above all else. 

The most suitable materials for the core magnet neutron shield are tungsten borides and metal hydrides \cite{brand:2025, windsor:2021} as they both offer superior flux attenuation. However, the physical isolation of the core magnet imposes extra constraints. Like the cryogenic region of the core magnet, there is no way to actively extract thermal energy from the shield during a fusion pulse. The only methods available in this scenario are radiating the thermal energy from the shield surface to the first wall, or storing it in an on-board reservoir to be extracted later at the end of the pulse. The internal space constraints make radiative cooling the preferable option as an on-board reservoir would need to store a significant portion of the plants output power and therefore require a significant amount of volume. Hence, the neutron shield requires materials with extremely high working temperatures exceeding $2000$~K in order to effectively reject the heat without placing severe constraints on the achievable fusion power density. Surface temperatures exceeding $2000$~K allow for wall loadings in excess of $1$~MW\;m$^{-2}$ which is required for any form of moderately compact, and therefore economically viable, fusion reactor. Increasing the temperature of the neutron shield beyond this point would allow for even higher wall loadings resulting in smaller an more attractive reactor designs. Therefore, the operating temperature of the neutron shield material is a key parameter that must be maximized in order to build economically attractive fusion power plants.

\begin{figure}[b!]
    \centering
    \includegraphics[width=1\linewidth]{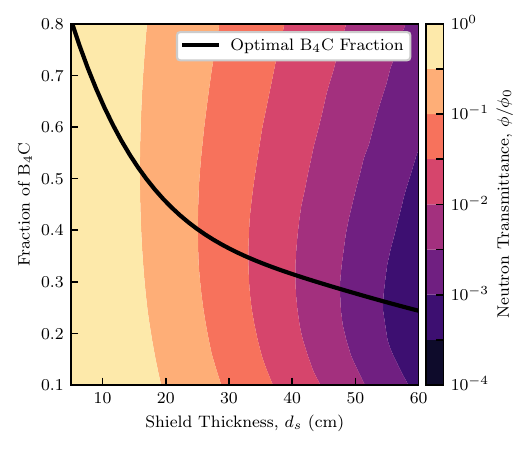}
    \caption{Plot of the neutron transmittance of a layered W-${\rm B}_4{\rm C}$-W neutron shield with varying ${\rm B}_4{\rm C}$ layer thicknesses. The absolute thickness of the optimal ${\rm B}_4{\rm C}$ layer remains relatively constant as the overall thickness of the shield changes.}
    \label{fig:boron_fraction}
\end{figure}

This extra requirement immediately rules out metal hydrides as they begin to decompose between $600-1000$~K \cite{hirooka:1984, pollard:2025}. This can be improved by forming a composite material with the metal hydride \cite{fletcher:2025}, however this still results in materials with operating temperatures below what is required here. Tungsten borides perform better as some phases are thermally stable up to  $2400$~K \cite{kvashnin:2018}, however they have yet to be manufactured at scale due to lower technological maturity. Therefore, for the purposes of this study we shall limit the neutron shielding to use well understood materials such as common tungsten alloys and boron carbide (${\rm B}_4{\rm C}$). If the use of these materials yields a neutron shield design that meets all requirements and results in an economically viable power plant, any advancements in material science will only act to improve performance and reduce the overall size of the reactor.

\begin{figure}[t!]
    \centering
    \includegraphics[width=1\linewidth]{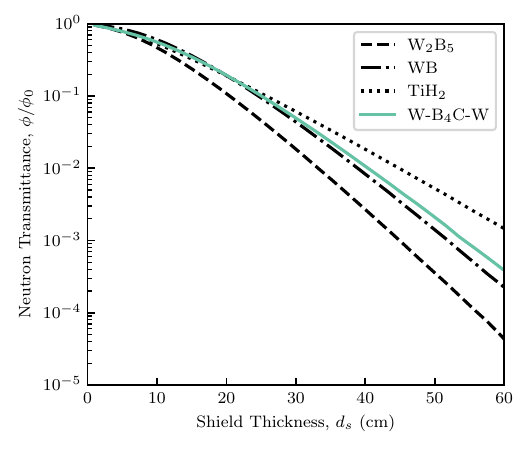}
    \caption{Comparison of the neutron attenuation performance of an optimal layered W-${\rm B}_4{\rm C}$-W shield and other commonly selected shielding materials.}
    \label{fig:shield_material}
\end{figure}

One advantage of pure tungsten is its extremely high melting temperature ($>3000\ ^\circ$C), albeit at the cost of reduced neutron attenuation performance compared to tungsten borides and metal hydrides. The performance of the overall shield can be improved back to tungsten boride levels by adding a layer of ${\rm B}_4{\rm C}$ to increase the rate of neutron absorption. The optimal thickness of this ${\rm B}_4{\rm C}$ layer, shown in Fig.~\ref{fig:boron_fraction}, was calculated for a range of shield thicknesses by solving a neutron transport problem on a thin column of material using the OpenMC code \cite{romano:2015}. The final performance of a layered shield with an optimal ${\rm B}_4{\rm C}$ fraction was then calculated to be similar to that of a monolithic block of mono tungsten boride (WB) as shown in Fig.~\ref{fig:shield_material}. 

It is also important for the neutron shield to maintain structural integrity during its operational lifetime. Here we will consider three main mechanisms that will lead to shield failure: tungsten recrystallization, thermal creep, and neutron damage. Tungsten undergoes recrystallization above temperatures of $\sim 1600$~K \cite{richou:2020, suslova:2014} which can make the material excessively brittle when cooled. The ductile-brittle transition temperature is a function of the level of recrystallization and is typically $<900$~K \cite{tietz:1965}. Therefore, if the shield is held above $900$~K, as it would be during standard plant operation, then the tungsten will only become brittle during infrequent plant maintenance shutdowns discussed in further detail in Section~\ref{sec:dp_RAMI}. 

The lifetime of the shield is therefore set by neutron damage and thermal creep effects. For the purposes of this study, it was assumed that the tungsten could withstand 1~MW-year/m$^2$ of neutron irradiation before it would require replacement \cite{NASEM:2021}. Thermal creep is then managed by first designating some portion of the shield to be held at a reduced temperature of approximately $600$~$^\circ$C (warm shield) through the use of a secondary thermal reservoir. This ensures part of the neutron shield remains structurally stiff to support the remaining region (hot shield) which is radiatively cooled and split into tiles, as shown in Fig.~\ref{fig:magXsection}, to reduce gravity induced stresses and hence also thermal creep. The natural transition point between the two regions is between the outer tungsten layer and the ${\rm B}_4{\rm C}$ layer as the majority of the neutron energy is deposited in the tungsten and ${\rm B}_4{\rm C}$ has a lower melting point. OpenMC models predict that for shields with relevant thicknesses, $\sim90$~\% of the incident neutron energy is deposited in the outer tungsten layer leaving the rest to be transferred to the secondary reservoir. The interface between the warm and hot shields (${\rm B}_4{\rm C}$ layer and tungsten tiles respectively) will need to have a low thermal conductivity to enable efficient heat rejection through radiation at the shield surface. For the purposes of this study, a thermal break with a $10$~W\,m$^{-2}$K$^{-1}$ conductance is assumed and will be treated as a requirement for future detailed shield designs. Further design of the neutron shield---including details such as shine through prevention, tile mounting mechanism, and specific secondary reservoir coolant choice---is an ongoing area of research.

The neutron energy deposited in the shield can be recovered. The energy radiated from the shield surface to the first wall will eventually be conducted to the tritium breeder blanket covered in Sections~\ref{sec:dp_vac} and \ref{sec:dp_tritium_breeding} where it will add to the total thermal output. The working fluid stored in the secondary reservoir will be removed when the magnet is docked where, due to its high temperature, it may be passed through an exchanger to extract useful energy. 

Finally, the end of life of the shield is considered here as it can have a significant effect on the overall cost of the plant. The main damage mechanisms described in this section result in effects which can be repaired through reprocessing of the shield material. The outer tungsten tiles, which will see the most rapid degradation, can be removed and directly recycled into new tiles for future core magnets after a cool down period of $\sim 1$~year due to their low activation half life and relatively low transmutation rate \cite{windsor:2022}. The end of life processes for the ${\rm B}_4{\rm C}$ layer is complicated by the production of tritium through the $^{10}$B(n, $\alpha$)$^3$H reaction. However, the ${\rm B}_4{\rm C}$ layer sees significantly lower fluences than the hot-shield tiles and is also significantly less expensive, which avoids the need to consider recycling the material. 

\subsubsection{Vacuum Vessel}\label{sec:dp_vac}

\begin{figure}[t!]
    \centering
    \includegraphics[width=1\linewidth]{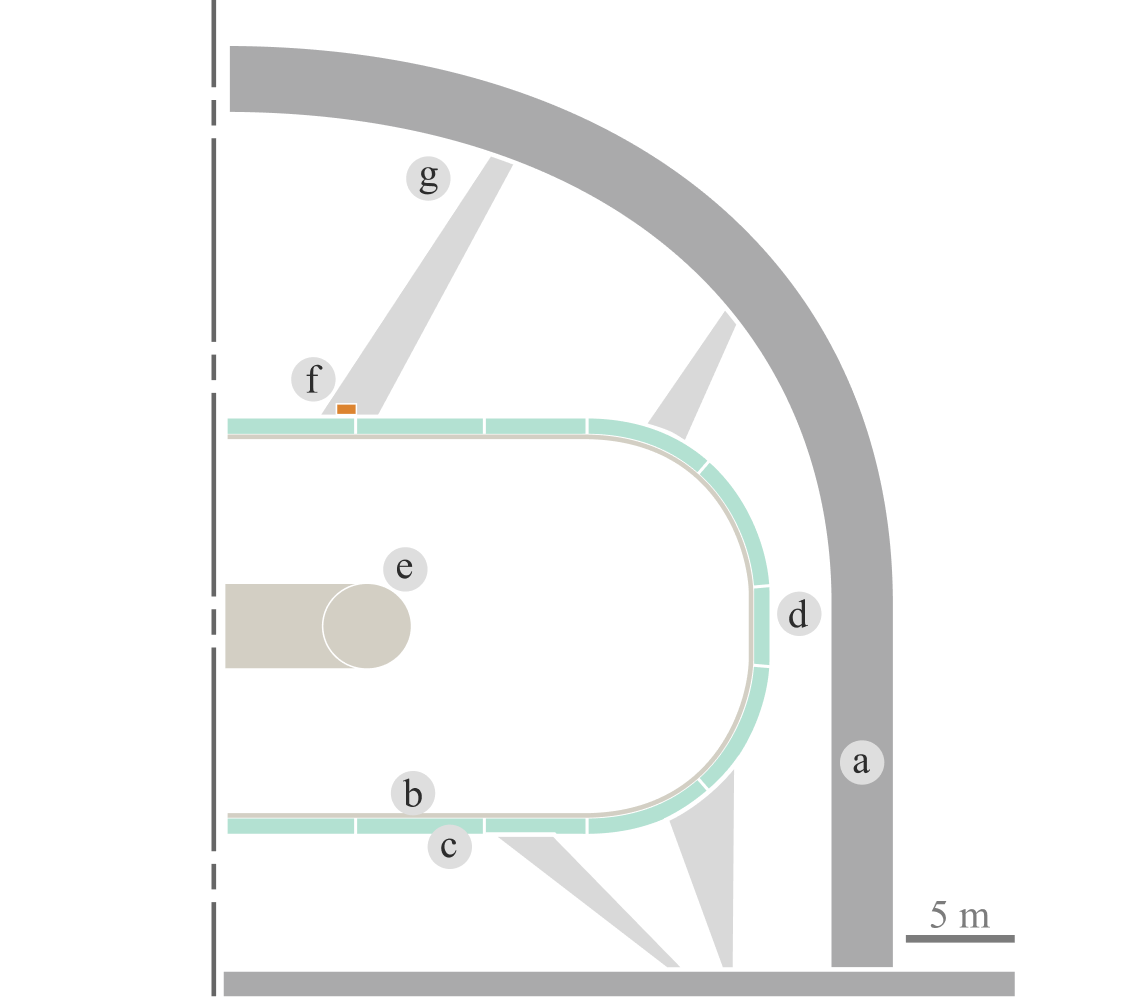}
    \caption{Representative cross section of a levitated dipole vacuum chamber, showing: (a) a reinforced concrete outer wall, (b) an inner, plasma facing wall, (c) the tritium breeding and thermal capture components, and (d) an annulus for plant services and maintenance access. The core magnet (e) is supported with a the top magnet (f). Structural members (g) transfer loads to the outer wall.}
    \label{fig:VacuumVesselSection}
\end{figure}

The vacuum vessels required for a levitated dipole fusion power plant are large due to the dependence described in Section~\ref{sec:dp_stability}, where peak pressures are related to radius as in Eq.~\eqref{eq:p_ratio}. However, a feature that is unique to this magnetic confinement scheme is that the vessel does not need to make accommodations for large high-field coils and their associated loads. Therefore, the vessel design can be simple and cost effective with a representative cross section shown in Fig.~\ref{fig:VacuumVesselSection}. The outer wall of the reactor is proposed to be a thick dome constructed from reinforced concrete. This outer wall provides a rough vacuum ($10$~Pa), and is designed to manage the loads from the vacuum pressure differential and the weight of the top and core magnets. Internal to the outer vessel, a tritium breeding blanket, as described in Section~\ref{sec:dp_tritium_breeding}, forms the bulk of the inner vacuum vessel. The plasma facing side of the inner vacuum vessel would be constructed from a thin layer of Inconel 718 with a surface coating of tungsten. Aside from tritium breeding, the purpose of the inner vacuum vessel is to provide the final high-vacuum conditions required for fusion. Since the outer vessel already provides a rough vacuum, the pressure differential over the inner vacuum vessel is small enough to only require adequate structure to support the mass of the tritium breeding blanket. Additionally, the loads created by the interaction between the top and core magnets will be transferred directly to the outer vacuum vessel.

At least one large opening is required to remove and install the core magnet and other components. Other penetrations include ports for plasma heating, fuel handling, cryogenic slush, and fluid transfer for heat extraction.  A reasonable analogue to this vessel is The Space Power Facility at NASA's Glenn Research Center \cite{sorge:2013}, which is similar in size and function. The design and build of such a vessel is therefore not anticipated to be a large technical risk. The decoupling of the inner and outer wall allow for large spaces to be added for maintenance, gantries and other services without incurring significant additional cost. An allowance has been made for a $2$~m wide cavity between the breeding blanket and the inside of the outer wall, allowing plentiful space for vessel maintenance operations. 

\subsubsection{Tritium Breeding}\label{sec:dp_tritium_breeding}

\begin{figure}[t!]
    \centering
    \includegraphics[width=1\linewidth]{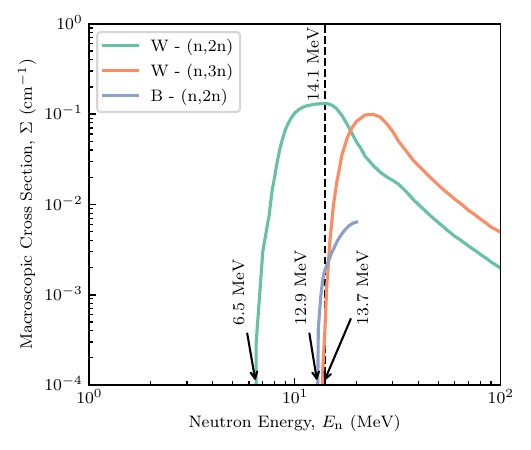}
    \caption{Macroscopic cross section of neutron multiplication reactions in the neutron shield calculated from the ENDF/B-VII.1 database \cite{chadwick:2011}. The outer tungsten layer will see significant numbers of both (n, 2n) and (n, 3n) reactions.}
    \label{fig:tungsten_multiplication}
\end{figure}

The reactors proposed in this study will use a DT fuel cycle which will require a tritium breeding blanket. They are designed to operate at a tritium breeding ratio (TBR) of 1.1 which is assumed to be sufficient for self fueling and supply for future reactors. There are theoretically two surfaces available for tritium breeding, the core magnet neutron shield and the first wall. However, all materials suited for tritium breeding do not have the neutron flux attenuation performance necessary for protecting the core magnet. Instead, the outer tungsten shield behaves as a neutron reflector due to it's high neutron multiplication cross section shown in Fig.~\ref{fig:tungsten_multiplication}, which accounts for the neutrons absorbed in the shield. The final neutron current that passes through the first wall will no longer be comprised solely of $14.1$~MeV neutrons, but instead the reflection process will introduce a significant population of $\sim 1$ MeV neutrons. These neutrons will not be able to undergo the endothermic $^7$Li(n, n'$\alpha$)$^3$H reaction which is responsible for giving materials TBRs greater than unity as it has a threshold energy of $2.5$~MeV. Initial TBR calculations show that the effect from this low energy neutron population are negligible, and hence will be treated that way for the remainder of this study. More in depth modeling and design of a tritium breeder blanket will be left as a topic for future study.

\begin{figure}[t!]
    \centering
    \includegraphics[width=1\linewidth]{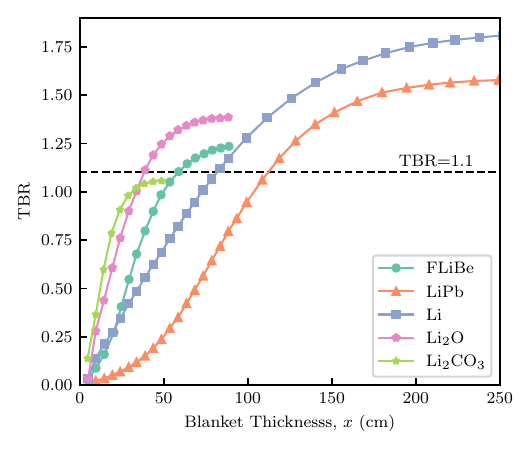}
    \caption{TBRs of varying thicknesses of materials \cite{shanliang:2003} relevant to levitated dipole reactors.}
    \label{fig:blanket_thickness}
\end{figure}

The simple vacuum vessel structure lends itself to thick breeder blankets capable of achieving high TBRs. Additionally, the magnetic field at the first wall is small and steady state allowing for liquid metal blanket materials to be used without the need to consider MHD effects. These benefits would theoretically free up the choice of blanket material, however, in reality the size of the inner vessel places constraints on the cost of blanket material per square meter. As shown in Fig.~\ref{fig:blanket_thickness}, materials such as LiPb and Li can achieve high TBRs with thick blankets, however the surface area these blankets need to cover would make the material costs prohibitive. Instead ceramic blanket materials show the most promise due to the lower blanket thickness \cite{shanliang:2003}, aiming to reduce overall system cost. Traditionally, ceramic blankets are discounted due to the increased maintenance requirements. This is not expected to be an issue for a levitated dipole due to the simple vacuum vessel structure allowing ample room for blanket access. This study assumes the use of a ${\rm Li_2 O}$ blanket as a performance benchmark. Other materials can match the TBR of ${\rm Li_2 O}$ with the help of a neutron multiplier, however the exact composition is not needed in order to determine the overall plant performance. 

\subsubsection{Plasma Heating Systems}\label{sec:dp_heating_systems}

Plasma heating in levitated dipoles can be accomplished using several established auxiliary heating technologies. Previous dipole experiments have primarily employed electron-cyclotron resonance heating (ECRH) \cite{garnier:2006, nishiura:2015} and ion-cyclotron resonance heating (ICRH) \cite{nishiura:2015a}, while early theoretical work by Hasegawa also proposed neutral beam injection (NBI) as a viable heating mechanism \cite{hasegawa:1990}.

ECRH has been successfully demonstrated on multiple dipole experiments and offers a relatively straightforward heating approach with favorable absorption characteristics. A particular advantage for dipole geometries is the ability to launch waves vertically into the high-field side, enabling the use of high-frequency gyrotrons with high cutoff densities. However, current gyrotron systems suffer from low wall-plug efficiency (typically 30-40\%) and rely on a specialized supply chain, which may present challenges for commercial deployment.

ICRH was demonstrated on the RT-1 experiment with mixed results. This approach benefits from higher efficiency RF sources compared to ECRH, approaching $70$~\% \cite{jardin:2006, faugel:2020}, and access to a more established industrial supply chain. The primary disadvantage lies in the increased complexity of antenna design and wave propagation modeling in the dipole magnetic geometry, which introduces greater scientific uncertainty in predicting heating performance. Investigation of ICRH in higher performance levitated dipole devices is currently ongoing \cite{wallace:2025}. 

NBI represents a lower-risk heating option with well-understood physics and mature technology. Unlike conventional magnetic-confinement fusion concepts, dipole reactors are less constrained by the large vessel penetrations required for neutral beam injection. Further, dipoles present improved penetration pathways to the plasma core and avoid the need for immature negative source ion beams even at reactor scale. Nevertheless, low wall plug efficiencies and specialized supply chain may prove prohibitive for power plant economics.

OpenStar's experimental program will systematically evaluate these heating methods in future devices to inform the selection of optimal heating systems for future commercial-scale implementations based on the trade-offs outlined above. The plasma parameters employed in the present modeling study assume ICRH as the baseline heating mechanism.

\subsection{Plant Operation}\label{sec:dp_RAMI}

Reliability, Accessibility, Maintainability, and Inspectability (RAMI) analysis for traditional fusion technologies often results in an emphasis on component lifetime due to the lengthy periods of downtime required for maintenance as the result of a comparative lack of accessibility. Levitated dipoles mitigate this by having excellent accessibility and maintainability due to the available space and modularity of the system, which takes considerable pressure off the overall RAMI requirements. Indeed, RAMI stands as one of the most challenging barriers to fusion energy deployment, and dipoles offer a unique pathway to an economic RAMI strategy.

\subsubsection{Magnet Replacement \& Maintenance}

The complete decoupling of the core magnet from the vacuum vessel allows for it to be treated as a semi-consumable item which is replaced once damage from the fusion neutrons degrades its performance below acceptable levels. As discussed in Section~\ref{sec:dp_hts_coil_design}, a portion of the coil is designated as ``sacrificial'' and will experience a higher neutron flux than the rest of the coil. Once the degradation limit in this region is reached, the entire core magnet can be removed from the chamber and replaced with a refurbished unit. The replacement process involves first discharging the core magnet which should be relatively quick, occurring in a matter of days, assuming a functioning quench protection system is already built into the magnet. Next, the magnet is removed through the bottom of the inner vacuum vessel and passed through an airlock of some sort into a hot cell external to the reactor. The airlock in this case is desirable as it prevents load cycling of the outer vacuum vessel while also cutting down on the time required to pump down to full vacuum. The new core magnet is then passed in and charged in place. This charging process is expected to be the lengthiest part of the replacement process, however, we still expect the total down time to be less than $2$~weeks. 

The damaged magnet---which is sitting in an external hot cell---can now be maintained without affecting plant down time, and in an environment without spatial constraints. The damaged sacrificial REBCO conductor is then replaced to create a refurbished core magnet that can be reused. We expect to be able to design the sacrificial section of the coil such that it comprises less than $20\%$ of the overall coil cross section, with the remainder lasting around ten times longer due to the shielding effects of the sacrificial region. Ideally, multiple reactors would share the same maintenance facility allowing for the more efficient use of replacement core magnets. For the purposes of this study however, we assume that each reactor will have an attached maintenance facility. In the case where the magnet needs to be replaced once a year, the power plant would be able to achieve an overall availability factor of more than $95\%$.

\subsubsection{Vessel Maintenance}

As discussed in Section~\ref{sec:dp_vac}, ample space can be allocated for access to the inner vacuum vessel without significantly affecting the cost of the overall reactor. This space allows easy access to the inner vacuum vessel and tritium breeding blanket with a relatively simple robotic system. Therefore, technologies that require more active and frequent maintenance become viable, reducing design constraints and thereby the overall cost of the plant. Furthermore, the longevity of the plasma facing components and blanket should be higher than in other confinement concepts because of the lower wall loading afforded by the large chamber size. The maintenance strategy for these components is once again aimed at reducing plant downtime, for which the best strategy is to take advantage of the system modularity and perform the maintenance while the core magnet is being swapped out.

\section{Design Methodology}\label{sec:methods}

To fully capture the interaction between the systems outlined in Section~\ref{sec:design_principles}, this study uses a full system optimization approach to generate reactor designs. The na\"ive choice of cost function for this problem would be the plant LCOE constrained with either plant size or total capital cost. However, for such a cost function to give valid results a method for calculating the energy confinement time would need to be provided. As outlined in Section~\ref{sec:dp_intro_scaling}, no such model exists for dipoles. Therefore, we take a reversed approach: instead of optimizing for reactor power under the constraints of expected device performance, we design a reactor assuming a value for $Q_{\rm sci}$ and design to minimize the required confinement time of a small scale demonstration device. In this way the design points produced by this process have the greatest chance of being viable. Using this process we design two reactors: Reactor A which aims to be feasible assuming conservative Bohm-like scaling, and a smaller Reactor B which targets a smaller overnight capital cost while requiring more aggressive performance targets.

\subsection{Cost Function}\label{sec:m_cost_function}

To minimize the required energy confinement time, $\tau_{\rm e}$, for a demonstration reactor, we first need to define the scaling to apply to $\tau_{\rm e}$ of the generated power plant reactor. For this purpose we use the following formulation \cite{waltz:1990}:
\begin{equation}
    \label{eq:taue_definition}
    \tau_{\rm e}=\frac{k_{\rm \alpha}}{\langle\Omega_{\rm i}\rho_*^\alpha\rangle_{\rm mid}},
\end{equation}
where $\rho_*=\rho_{\rm i} / a$, is the relative Larmor radius normalized to the plasma minor radius, $a=R_{\rm lcfs}-R_0$. Due to the highly peaked nature and shape of the plasma, $\rho_*$ cannot be assumed to be constant as it is in a tokamak and is treated here as a function of both $r$ and $z$. The remaining parameters are the ion gyrofrequency, $\Omega_{\rm i}$; a constant of proportionality that captures the confinement performance, $k_{\rm \alpha}$; and $\alpha$ which represents either Bohm-like ($\alpha=2$) or gyro-Bohm-like ($\alpha=3$) scaling. We have also chosen to take the average over the low field side midplane for $\langle\Omega_{\rm i}\rho_*^\alpha\rangle_{\rm mid}$ to account for the significant non-uniformity of the plasma profile, as discussed further in \ref{app:geometry_factors}. We then reformulate Eq.~\eqref{eq:taue_definition} to extract a ``device index'' $\xi_{\alpha}$:
\begin{equation}
    \label{eq:device_index}
    \tau_{\rm e}=k_{\rm \alpha}\left\langle\frac{T^{\alpha/2}}{\xi_{\alpha}}\right\rangle_{\rm mid}^{-1} \ \text{with}\ \ \xi_{\alpha}=a^\alpha B^{\alpha-1},
\end{equation}
allowing the separation of the physical embodiment of a reactor, represented by $\xi_{\alpha}$, from the operating point defined by the ion temperature $T$ at constant pressure and $\beta_0$. 

The goal of minimizing the required $\tau_{\rm e}$ of a small scale demonstration device can therefore be achieved by minimizing the implied $k_{\rm \alpha}$ of a reactor while assuming a fixed $Q_{\rm sci}$. This is distinct from minimizing $\tau_{\rm e}$ as by fixing $Q_{\rm sci}$ we have restricted any two reactors that produce the same fusion power and have the same plasma stored energy to also have the same $\tau_{\rm e}$ even if they have different values of $\xi_{\alpha}$. Additionally, the demonstration device will operate with a set $\xi_{\alpha}'<<\xi_{\alpha}$ and plasma operating point determined by what is financially reasonable. Therefore, in accordance with Eq.~\eqref{eq:device_index} the only factor that impacts $\tau_{\rm e}$ of the smaller device is the value of $k_{\rm \alpha}$. 

The implied $k_{\rm \alpha}$ of a reactor is calculated from Eq.~\eqref{eq:taue_definition} utilizing the 0D power balance:
\begin{equation}
    \label{eq:tau_e_estimate}
    \tau_{\rm e} = \frac{U_{\rm p}}{f_{\rm sh}f_{\alpha}P_{\rm fus} + P_{\rm aux} - P_{\rm rad}},
\end{equation}
where $f_{\rm sh}$ is the fraction of $\alpha$ particle energy that contributes to self heating, $f_{\alpha}$ is the fraction of the fusion reaction energy released as an $\alpha$ particle, and $P_{\rm rad}$ is the total power lost from the plasma in the form of radiation. As in Section~\ref{sec:dp_peak_location}, it has been assumed that the $\alpha$ power deposited in $\mathfrak{R}_{\rm GC}$ will be entirely balanced by radiation losses. The total radiated power is then modeled as the sum of the bremsstrahlung power, the main radiative mechanism, and the good curvature region $\alpha$ heating power:
\begin{equation}
    P_{\rm rad} =  P_{\rm brem} + f_{\rm gcr}f_{\alpha}P_{\rm fus}.
\end{equation}
Defining the exact form of this additional radiative term is outside the scope of this study. If the self heating fraction is then defined as $f_{\rm sh}=f_{\rm gcr}+f_{\rm bcr}$, the contribution of the $\alpha$ heating in $\mathfrak{R}_{\rm GC}$ cancels and we are left with:
\begin{equation}
    \label{eq:taue_from_q}
    \tau_{\rm e} = \frac{\frac{3}{2}(1+\bar{Z})\langle nT\rangle}{\frac{E_{\rm f}}{4}\left(f_{\rm bcr}f_{\alpha}+\frac{\eta_{\rm h}}{Q_{\rm sci}}\right)\left\langle n^2\langle\sigma v\rangle_{v}\right\rangle - C_{\rm B}Z_{\rm eff}\bar{Z}^2\left\langle n^2T^{1/2}\right\rangle}, 
\end{equation}
where for the sake of this optimization we have used $\bar{Z}\equiv n_{\rm e}/n_{\rm i}=1.2$ and $Z_{\rm eff}\equiv \sum_i n_i Z^2_i/n_{\rm e}=1.5$ to calculate the bremsstrahlung losses.

However, simply minimizing $k_{\rm \alpha}$ by itself does not define a well-posed optimization problem. According to Eq.~\eqref{eq:device_index} the size of the device, represented by $\xi_{\alpha}$, is inversely proportional to $k_{\rm \alpha}$ and therefore a global minimum cannot be defined. The optimizer will act to minimize $k_{\rm \alpha}$ by increasing $\xi_{\alpha}$ to arbitrary levels. If we assume that the lower bound on overnight capital cost of the plant is a monotonically increasing function of the device size captured in $\xi_{\alpha}$, then this would also result in arbitrarily high plant costs. This is prevented by setting a limit on the capital cost which in turn defines an optimal device size, $\xi_{\alpha}^*$, which has the lowest achievable $k_{\rm \alpha}$ allowed by the cost constraint. Therefore, this cost function requires a constraint on the overnight capital cost of the plant in order to converge.

The exact optimal point is determined by whether the energy confinement time scales in a Bohm-like or gyro-Bohm like fashion. This study focuses on a Bohm-like scaling as it is the more conservative of the two, and hence reactors that are feasible given this scaling represent conservative design points. In the case where a future dipole displays better scaling, it would then be possible to design smaller and more capital-efficient reactors than those presented in this paper. Therefore, the cost function used in the optimization process is explicitly:
\begin{equation}
    \label{eq:kbohm}
    \min\left(\tau_{\rm e}\left\langle\frac{T}{a^2B}\right\rangle_{\rm mid}\right),
\end{equation}
with a constraint placed on the total overnight capital cost of the implied power plant.

\subsection{Optimization Process}\label{sec:m_integrated_modeling}

\begin{table}[b!]
    \centering
    \resizebox{\columnwidth}{!}{
    \begin{tabular}{l l l l}
        \hline\hline
        Parameter & Symbol & Range & Units \rule{0pt}{4ex}\\ [1.5ex]
        \hline
        REBCO Coil Outer Radius & $R_{\rm c}$ & $[0,\ 10]$ & m \rule{0pt}{2.6ex}\\ [0.5ex]
        Magnet Shape Control Points & $[r_1,\ z_1,\  ...\ ,\ r_5,\ z_5]$ & $[0,\ 1]$ & $R_{\rm m}$ \\ [0.5ex]
        Operating Current Density & $J_{\rm op}$ & $[0,\ 300]$ & A/mm$^2$ \\ [0.5ex]
        Cryogen Reservoir Thickness & $d_{\rm r, cryo}$ & $[0,\ 1]$ & m \\ [0.5ex]
        First Wall Radius & $R_{\rm fw}$ & $[1.1,\ 20]$ & $R_{\rm m}$ \\ [0.5ex]
        Pressure Peak Location & $\psi_0$ & $[0,\ 0.5]$ & \\ [0.5ex]
        Core Temperature & $T_0$ & $[1,\ 30]$ & keV \\ [0.5ex]
        Edge Pressure & $p_{\rm lcfs}$ & $[0,\ 1000]$ & Pa \\ [0.5ex]
        Neutron Shield Limiting Thickness & $d_{\rm ns}$ & $[0,\ 1]$ & m \\ [0.5ex]
        Neutron Shield Reservoir Thickness & $d_{\rm r, ns}$ & $[0,\ 1]$ & m \\ [0.5ex]
        \hline\hline
    \end{tabular}
    }
    \caption{Reactor optimization variables and their respective allowed ranges.}
    \label{tab:ReactorVariables}
\end{table}

\begin{figure*}[t!]
    \centering
    \includegraphics[width=1\linewidth]{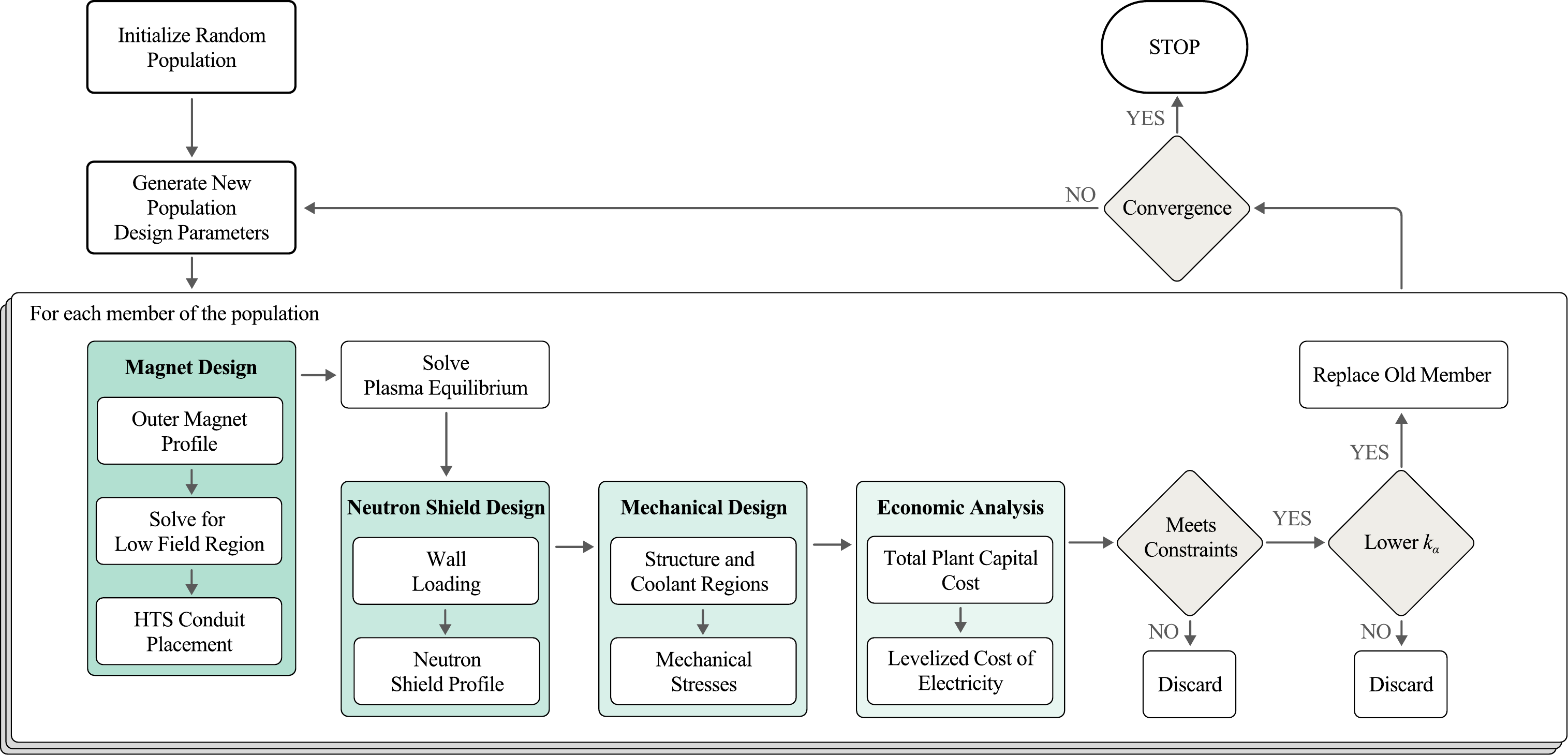}
    \caption{Flow chart of the reactor optimization process. A reactor is described using the set of variables given in Table~\ref{tab:ReactorVariables} and progressively built up using the shown modeling tool-chain.}
    \label{fig:optimization_flow}
\end{figure*}

Designing a reactor that globally minimizes Eq.~\eqref{eq:kbohm} while still satisfying engineering and economic constraints requires the modeling of the full plant. The tightly coupled nature of the core magnet performance, core magnet geometry, and the final performance of the reactor require this modeling to take place within the optimization loop. This study uses the parameterization summarized in Table~\ref{tab:ReactorVariables} which, once constraints are applied to the magnet shape parameters, defines a 14 dimensional design space which is searched to find the global minimum. The sensitivity of the final reactor performance to some of these parameters rules out a brute force search as a viable method for finding the global optimum. Additionally, the high level nature of the constraints and the complexity introduced by the interacting systems also makes gradient- based optimization methods unsuited to this task. Instead, differential evolution \cite{storn:1997} (DE) was chosen for this study as it offers good flexibility and adequate performance while simultaneously remaining easy to implement, allowing more attention to be focused on the modeling stages that comprise the optimization loop as outlined in Fig.~\ref{fig:optimization_flow}. The parameters in Table~\ref{tab:ReactorVariables} are generated using DE and then passed through a series of modeling stages that progressively build up the reactor from the REBCO coil outwards.

\subsubsection{REBCO Coil Design}\label{sec:m_hts_coil_optimisation}

The first stage in the optimization process is the design of the REBCO coil that generates the confining magnetic field. This process is complicated by the requirement of a low-field region within the magnet to accommodate for the on-board superconducting power supply and electronics as introduced in Section~\ref{sec:dp_fp}. First, the upper half of the outer shape of the coil is constructed using a fourth order B\'{e}zier curve which is constrained to have the required outer radius and be vertical at the mid plane to ensure continuity. The resulting parameterization has five degrees of freedom that are determined through the optimization. Then, the boundary of the low-field region is created and optimized to reduce the interior magnet field strength.

The homogeneous current density is then divided into a grid of REBCO conductor conduits to represent the final winding geometry. The homogeneous region is first broken into cells whose required cross sectional area is constrained by the ratio of the target operating current, $I_{\rm op}$, to $J_{\rm op}$. The cell width is set by the turn with the lowest expected tape critical current, $I_{\rm c}$, which will require the highest number of parallel tapes to carry the target $I_{\rm op}$. The REBCO tape $I_{\rm c}$ is most sensitive to magnetic field perpendicular to the tape's surface, hence the lower $I_{\rm c}$ cells will be located at the top and bottom of the coil. All other cells in the coil will require fewer parallel tapes, and therefore can support a higher ratio of structural material helping with transferring the Lorentz load to the structural over-band. Therefore, the coil can be broken into regions of high conductor-ratio conduit and high steel-ratio conduit to optimize the performance and mechanical strength of the coil. The design of the conduit beyond this point is outside the scope of this study and will be a topic of future publications.

The REBCO tape performance is based off the performance of SuperOx YBCO presented in the Robinson Supercurrent Database \cite{wimbush:2021} and extrapolated to the required operating conditions using an elliptic function \cite{grilli:2014}. Recently, Faraday Factory, a REBCO tape manufacturer, announced a new ``Mirai'' family of REBCO tape which is expected to reliably produce engineering current densities in excess of $1000$~A\,mm$^{-2}$. This corresponds to a performance increase of $\sim30$~\% over their current generation product \cite{wimbush:2021, molodyk:2021}. This improvement is applied on top of the modeled tape $I_{\rm c}$ extrapolated from the available data. The final grid of REBCO conduit will affect the optimization of the low-field region from the previous step. However, any deviation from this optimum is assumed to be correctable with the use of passive and active magnetic shielding and therefore ignored for the purposes of this optimization. The final output from this model is a grid of currents each with a required number of parallel tapes needed to carry the target $I_{\rm op}$.

\subsubsection{Plasma Equilibrium}\label{sec:m_plasma_equilibrium}

The core magnet coil is then used as an input to the DipolEQ MHD equilibrium code \cite{garnier:1999} which solves the reduced (toroidal-field-free) Grad-Shafranov equation:
\begin{equation}
    \nabla \cdot \left( \frac{\nabla \Psi}{R^2} \right) =
    -\mu_0 \frac{\d p}{\d\Psi},
    \label{eq:GS}
\end{equation}
in the region between the first and last closed flux surfaces. To define the pressure, density, and temperature profiles a normalized poloidal magnetic flux is defined:
\begin{equation}
    \label{eq:normpsi}
    \psi \equiv \frac{\Psi - \Psi_{\rm fcfs}}{\Psi_{\rm lcfs} - \Psi_{\rm fcfs}} \in [0,\ 1].
\end{equation}
The choice of pressure profile used in Eq.~\eqref{eq:GS}, assumed to be isotropic for simplicity, is based on whether the plasma is in a region of good or bad curvature. In the bad curvature region the pressure follows Eq.~\eqref{eq:p_ratio} and in the good curvature region the pressure profile is arbitrarily defined as a cosine function:
\begin{equation}
    \label{eq:pressure_profile}
    p(\psi) = 
    \begin{cases}
      \frac{p(\psi_0)}{2}\left[1-\cos{\left(\pi\frac{\psi}{\psi_0}\right)}\right] & \text{in } \mathfrak{R}_{\rm GC}\ (0 \leq \psi< \psi_0) , \\
      p_{\rm lcfs}\left(\frac{\delta V_{\rm lcfs}}{\delta V(\psi)}\right)^\gamma  & \text{in } \mathfrak{R}_{\rm BC}\  (\psi_0\leq\psi<1),
    \end{cases}
\end{equation}
where the peak location, $\psi_0$, and an edge pressure, $p_{\rm lcfs}$, need to be provided by the optimizer. 

The density and temperature profiles can be obtained by realizing that Eq.~\eqref{eq:pvgamma} can be equivalently expressed as $\delta(n\delta V)=0$ and $\delta\left(T\delta V^{\gamma-1}\right)=0$. These are then combined and integrated to obtain an explicit form of the density profile:
\begin{equation}
    \label{eq:density_profile}
    n(\psi)=\left(\frac{p(\psi)}{p'}\right)^{\frac{1}{\gamma}}n',
\end{equation}
and temperature profile:
\begin{equation}
    \label{eq:temperature_profile}
    T(\psi)=\left(\frac{n(\psi)}{n'}\right)^{\eta}T'.
\end{equation}
where the primed variables represent reference values, such as at the plasma edge or the pressure peak. The optimization algorithm provides the core temperature, $T_0$, to enable this calculation. Using these profiles, the fusion and neutron production rates are calculated and then interpolated onto an $(r,\ z)$ grid to be used by the following models.

The optimizer is given control of $\psi_0$ in order to account for the large gyro radius of the 3.5 MeV $\alpha$ particles. To maximize plasma performance a levitated dipole reactor design should always aim to place $\Psi_0$ as close to $\Psi_{\rm fcfs}$ as possible. However, at smaller values of $\Psi_{\rm fcfs}$ the prompt loss of $\alpha$ particles to the core magnet will increase to unacceptable levels. To prevent this, a constraint is placed on the minimum real-space separation between $\Psi_{\rm fcfs}$ and $\Psi_0$ expressed as a multiple of the $\alpha$ particle gyro-orbit. This separation is impossible to calculate before producing the equilibrium, therefore it is necessary to allow the optimizer to control the peak location to iteratively approach the correct location.

\subsubsection{Neutron Shielding Design}\label{sec:m_neutron_shield_optimisation}

The next stage in the optimization process is the design of the neutron shield. As discussed in Section~\ref{sec:dp_Neutron_Shield}, all space up to $\Psi_{\rm fcfs}$ will be used to maximize shield thickness without encroaching on the region of closed field lines. The wall loading neutron flux can be calculated at points lying on $\Psi_{\rm fcfs}$ by integrating over the plasma volume accessible via straight-line trajectories:
\begin{equation}
\Phi_{\rm{flux}}(\vec{r}) = \int_{V_{\rm{visible}}} \frac{\rho_{\rm{rate}}(\vec{r}\,')}{4 \pi |\vec{r} - \vec{r}\,'|^{2}} \hat{n}(\vec{r}) \cdot \frac{(\vec{r}\,'-\vec{r})}{|\vec{r} - \vec{r}\,'|}\ \text{d}\vec{r}\,',
\label{eq:wall-loading}
\end{equation}
where $\hat{n}$ and $\rho_{\rm{rate}}$ are the the unit surface normal and reaction rate density, respectively. Projection onto the $rz$ plane reduces the numerical evaluation of Eq.~(\ref{eq:wall-loading}) to an effective 2D problem, for which polynomial straight-line intersections can be determined analytically. 

The shield is then defined from the closest point on $\Psi_{\rm fcfs}$ to the core magnet coil. Here the wall loading neutron flux is sampled and used alongside the optimizer determined characteristic shield thickness, $d_{\rm ns}$, to calculate the shielded flux using the attenuation data for a W-${\rm B}_4{\rm C}$-W composite structure as presented in Fig.~\ref{fig:shield_material}. This predicted neutron flux is then used to calculate the neutron shield thickness around the rest of the core magnet. 

This method shows good heating power correlation with the more accurate OpenMC models, however, it tends to underestimate the neutron flux in the shielded region. Contributions from geometric effects and material interactions, such as neutron multiplication, that are not included in this simple model account for this additional flux. In this study the lifetime of the REBCO tape used in the core magnet coil is calculated using a maximum fast neutron fluence of $3\times10^{18}$~cm$^{-2}$ \cite{fischer:2018}. As long as the neutrons can still be considered fast (energy $>0.1$ MeV), this limit is assumed to be agnostic of the incident neutron direction. Therefore, the additional neutron flux modeled in OpenMC is assumed to impact the lifetime of the tape. To account for this mismatch, the flux predicted in this simplified model is multiplied by a factor of 3 informed from numerous comparisons with OpenMC. 

\subsubsection{Internal Structure Design}\label{sec:m_structure_optimisation}

Once the core magnet coil and neutron shield have been defined, the remaining volume can then be distributed among the core magnet coil structure, cryogenic reservoir, and the neutron shield reservoir. In order to properly capture the design pressures acting on each of these components, the optimizer is given control of the volume of both the cryogen and shield reservoirs through their midplane thicknesses, $d_{\rm r, cryo}$ and $d_{\rm r, ns}$ respectively. The remaining volume is then allocated to the core magnet coil structural support. A more detailed design of the core magnet will alter the shapes of the reservoir to optimize the cooling power and docking speed. The geometry described here simply ensures there is enough volume of coolant available in the core magnet.

The stresses in the core magnet coil and structure are calculated using a 2D axi-symmetric FEA model. The grid of currents and magnetic field from the core magnet coil design are used to calculate the Lorentz body loads on the whole structure. For the optimization process it was sufficient to approximate the core magnet coil and structure as a homogeneous material. The calculated peak von Mises stress is then used to constrain the optimization process to ensure the final magnet design remains feasible.

\subsubsection{Reactor Performance}\label{sec:m_performance_optimisation}

\begin{table}[t!]
    \centering
    \resizebox{\columnwidth}{!}{
    \begin{tabular}{l l l l}
        \hline\hline
        Parameter & Symbol & Value & Units \rule{0pt}{4ex}\\ [1.5ex]
        \hline
        Thermal Efficiency & $\eta_{\rm th}$ & $0.4$ & \rule{0pt}{2.6ex}\\ [0.5ex]
        Auxiliary Heating Efficiency & $\eta_{\rm aux}$ & $0.7$ &  \\ [0.5ex]
        Cryogenic Efficiency & $\eta_{\rm cryo}$ & $0.0125$ &  \\ [0.5ex]
        Blanket Power Fraction & $f_{\rm b}$ & $0.8$ &  \\ [0.5ex]
        Neutron Energy Multiplier & $\zeta_{\rm ns}$ & $-0.12$ & \\ [0.5ex]
        Blanket Energy Multiplier & $\zeta_{\rm b}$ & $0.1$ & \\ [0.5ex]
        Electrical Heating & $P_{\rm \Omega}$ & $1.3$ & kW \\ [0.5ex]
        Shield Conduction Heating & $P_{\rm \kappa, cryo}$ & $1$ & kW \\ [0.5ex]
        Core Magnet Docked Time & $t_{\rm dock}$ & $5$ & min \\ [0.5ex]
        \hline\hline
    \end{tabular}
    }
    \caption{Parameters used for modeling of the thermal and net electrical power output of the reactors in this study.}
    \label{tab:ReactorPowerBalanceParameters}
\end{table}

The final stage in the optimization process is to calculate the net electrical output power of the plant, where the assumed input parameters have been summarized in Table~\ref{tab:ReactorPowerBalanceParameters}. The total thermal power is approximated by:
\begin{equation}
    P_{\rm th}= P_{\rm fus} + P_{\rm aux} + \zeta_{\rm ns}(1-f_{\rm b})P_{\rm n} + \zeta_{\rm b}f_{\rm b}P_{\rm n},
\end{equation}
where $\zeta_{\rm ns}$ and $\zeta_{\rm b}$ are the energy multiplication factors for the shield and blanket respectively; $f_{\rm b}$ is the fraction of neutron power deposited in the blanket; and $P_{\rm fus}$, $P_{\rm aux}$, and $P_{\rm n}$ are the fusion, auxiliary heating, and total neutron power respectively. For the purposes of this optimization process, the values are assumed to be independent of the plant geometry. As discussed in Section~\ref{sec:dp_tritium_breeding}, the tungsten layers of the neutron shield have a high cross section for endothermic neutron multiplication reactions. From OpenMC calculations on a range of dipole designs, the effect of this reaction is to reduce the heating in the shield by approximately 12~\%. The same OpenMC models were used to determine that approximately 80~\% of the neutron energy is deposited in the blanket. Meanwhile, the Li$_2$O breeder blanket is modeled as multiplying the incident neutron energy by $10$~\% \cite{sawan:2006}. The net electrical power is then modeled as:
\begin{equation}
    \label{eq:net_powaaahhh}
    P_{\rm net}= f_{\rm d}\left[\eta_{\rm th}P_{\rm th} - \frac{P_{\rm aux}}{\eta_{\rm aux}} - \frac{P_{\rm n, cryo} + P_{\rm \kappa, cryo} + P_{\rm \Omega}}{\eta_{\rm cryo}}\right],
\end{equation}
where $f_{\rm d}$ is the core magnet duty cycle, $\eta_{\rm th}$ is the electrical conversion efficiency, $\eta_{\rm aux}$ is the total efficiency of the auxiliary heating system, and $\eta_{\rm cryo}$ is the efficiency of the cryogenic cooling system (see Section~\ref{sec:dp_cryogen}) which we expect to be $1.25$~\%. The total cryogenic heat load is modeled as the sum of the cryogenic neutron and photon heating, $P_{\rm n, cryo}$, the heat conducted from the neutron shield, $P_{\rm \kappa, cryo}$, and the electrical resistive heating, $P_{\rm \Omega}$, required to keep the magnet charged. $P_{\rm n, cryo}$ is given by the previously described neutron shielding model in Section~\ref{sec:m_neutron_shield_optimisation}, whereas the values for $P_{\rm \Omega}$ and $P_{\rm \kappa, cryo}$ are given as a constant budget that must be met by later detailed engineering designs.

\begin{table}[b!]
    \centering
    \resizebox{\columnwidth}{!}{
    \begin{tabular}{l l l l}
        \hline\hline
        Parameter & Symbol & Value & Units \rule{0pt}{4ex}\\ [1.5ex]
        \hline
        Solid fraction & $f_{\rm solid}$ & $0.6$ & \rule{0pt}{2.6ex}\\ [0.5ex]
        \hline
        \textbf{Cryogenic Reservoir} \rule{0pt}{2.6ex}\\ [0.5ex]
        \hline
        Solid Density & $\rho_{\rm cryo}$ & $1.44$ & g cm$^{-3}$ \rule{0pt}{2.6ex}\\ [0.5ex]
        Latent Heat of Fusion & $L_{\rm cryo}$ & $16.6$ & J g$^{-1}$ \\ [0.5ex]
        Melting Temperature & $T_{\rm cryo}$ & $24.6$ & K \\ [0.5ex]
        \hline
        \textbf{Neutron Shield Reservoir} \rule{0pt}{2.6ex}\\ [0.5ex]
        \hline
        Solid Density & $\rho_{\rm ns}$ & $5.56$ & g cm$^{-3}$ \rule{0pt}{2.6ex}\\ [0.5ex]
        Latent Heat of Fusion & $L_{\rm ns}$ & $406$ & J g$^{-1}$\\ [0.5ex]
        Melting Temperature & $T_{\rm ns}$ & $571$ & $^\circ$C\\ [0.5ex]
        Neutron Power Fraction & $f_{\rm r}$ & $0.1$ &  \\ [0.5ex]
        \hline\hline
    \end{tabular}
    }
    \caption{Properties of the on-board cryogenic and neutron shield thermal reservoir materials. The cryogenic region uses a neon slush as the coolant \cite{ekin:2006} while the shield reservoir uses an aluminum-copper alloy \cite{shamberger:2020}.}
    \label{tab:CoolantProperties}
\end{table}

In order to model the core magnet duty cycle both the levitation time, $t_{\rm lev}$, and the docked time, $t_{\rm dock}$, are needed: 
\begin{equation}
    f_{\rm d}=\frac{t_{\rm lev}}{t_{\rm lev}+t_{\rm dock}}.
\end{equation}
This study assumes that the use of the latent heat based slush cryogen described in Section~\ref{sec:dp_cryogen} will allow short docked times of $t_{\rm dock}=5\ \text{min}$. The levitation time, on the other hand, is dependent on the volume of both reservoirs on-board the core magnet and the total amount of heating in those reservoirs. The total energy storage capacity of reservoir $i$ is calculated using:
\begin{equation}
    U_i = f_{\rm solid} V_i \rho_i L_i,
\end{equation}
Where $f_{\rm solid}$ is the fraction of solid material in the reservoir which is assumed to be 60\%, $V_i$ is the reservoir volume, $\rho_i$ is the coolant solid density, and $L_i$ is the coolant latent heat of fusion. The values for these parameters for both reservoirs are given in Table~\ref{tab:CoolantProperties}. The cryogenic reservoir limited levitation time can then be calculated as:
\begin{equation}\label{eq:tlevcryo}
    t_{\rm lev}^{\rm cryo} = \frac{U_{\rm cryo}}{P_{\rm n, cryo} + P_{\rm \kappa, cryo} + P_{\rm \Omega}},
\end{equation}
and the shield reservoir limited time as:
\begin{equation}\label{eq:tlevshield}
    t_{\rm lev}^{\rm shield} = \frac{U_{\rm shield}}{f_{\rm r}(1-f_{\rm b})P_{\rm n}},
\end{equation}
where $f_{\rm r}$ represents the fraction of shield neutron power that will be stored in an on-board reservoir which uses the latent heat of fusion of an aluminum-copper alloy \cite{shamberger:2020} to store the thermal energy until the next docking cycle. This material was selected to give a representative estimate of the total levitation time, the final material selection will be a topic of future study. The total levitation time, $t_{\rm lev}$, is then modeled as the minimum of Eqs.~\eqref{eq:tlevcryo} and \eqref{eq:tlevshield}:
\begin{equation}
    t_{\rm lev} = \text{min}\left(t_{\rm lev}^{\rm cryo}, t_{\rm lev}^{\rm shield}\right).
\end{equation}

\subsection{Optimization Constraints}\label{sec:m_performance_constraints}

\begin{table}[t!]
    \centering
    \resizebox{\columnwidth}{!}{
    \begin{tabular}{l l c c l}
        \hline\hline
        Name & Symbol & Reactor A & Reactor B & Units \rule{0pt}{4ex}\\ [1.5ex]
        \hline
        Relative Max Overnight Cost & & $1$ & $0.5$ & \rule{0pt}{2.6ex}\\ [0.5ex]
        Relative Max LCOE & & \multicolumn{2}{c}{1} & \\ [0.5ex]
        Target Q & $Q_{\rm sci}$ & \multicolumn{2}{c}{15} & \\ [0.5ex]
        Tritium Breeding Ratio &  & \multicolumn{2}{c}{1.1} & \\ [0.5ex]
        Max Edge Temperature & $T_{\rm lcfs}$ & \multicolumn{2}{c}{800}& eV \\ [0.5ex]
        Max Edge Pressure & $p_{\rm lcfs}$ & \multicolumn{2}{c}{$10^3$} & Pa \\ [0.5ex]
        Min $|\Psi_0 - \Psi_{\rm fcfs}|$ Separation &  & \multicolumn{2}{c}{$2$} & $d_{\rm l, \alpha}$ \\ [0.5ex]
        Max Von Mises Stress & $\sigma_\text{vm}$ & \multicolumn{2}{c}{700} & MPa \\ [0.5ex]
        Max Coil REBCO Fill Fraction &  & \multicolumn{2}{c}{40} & \% \\ [0.5ex]
        Low-Field Region Width & $d_{\rm lfr}$ & \multicolumn{2}{c}{150} & mm \\ [0.5ex]
        Max Neutron Shield Temperature &  & \multicolumn{2}{c}{2500} & K \\ [0.5ex]
        Min Sacrificial REBCO Lifetime &  & \multicolumn{2}{c}{1} & yr \\ [0.5ex]
        \hline\hline
    \end{tabular}
    }
    \caption{Design constraints applied to the reactor optimization process.}
    \label{tab:ReactorConstraints}
\end{table}

Constraints on the performance and total cost of the reactor, as summarized in table \ref{tab:ReactorConstraints}, are required to ensure the final power plant is economically viable and physically feasible. Each reactor is constrained with two economic parameters: overnight capital cost and levelized cost of electricity (LCOE). As discussed in Section \ref{sec:m_cost_function}, the constraint on the maximum overnight capital cost is core to the definition of the optimization problem as it allows for a global minimum to be defined. The constraint on the LCOE, on the other hand, acts to ensure the final reactor remains economically viable. OpenStar is currently in the process of developing a model for estimating the overnight capital cost and LCOE for levitated dipole fusion power plants which will be the topic of future work. This study uses preliminary results from this model which are subject to change as the model is developed. For this reason we avoid quoting specific values here, instead opting to present the relative cost and LCOE. Both reactors have been set with the same limit on the LCOE, but Reactor B was constrained to be less than half the overnight capital cost of Reactor A to encourage the design of a smaller plant. 

The remaining constraints were then imposed in an attempt to ensure the practical viability of the plant. The overall plant is assumed to have $Q_{\rm sci}=15$ and a $\mathrm{TBR}=1.1$ to be comparable with other proposed FOAK fusion power plants \cite{sorbom:2015, MANTA:2024, lion:2025}. The lifetime of the sacrificial portion of the core magnet coil, as discussed in Section~\ref{sec:dp_hts_coil_design}, is limited to be above $1$~year. This ensures the core magnet can be replaced during a planned yearly maintenance window, which is already accounted for in the economics of the plant. 

As mentioned in Section~\ref{sec:dp_edge}, the physics defining viable conditions at the plasma edge is not well understood. We expect there to be an edge pedestal due to preferential ion scrape off, however the magnitude of this effect is still unknown. I-mode tokamaks experience edge pedestals with temperatures and pressures exceeding $800$~eV and $10^3$~Pa respectively \cite{whyte:2010}, which we shall treat as an upper bound on performance in this study. The location of the pressure peak is constrained to reduce the possibility of excess prompt $\alpha$ losses. Scattering events with the background plasma will cause some of the $\alpha$ particles to transport further towards the first closed flux surface than what can be calculated using a simple orbit calculation. To account for this, $\psi_0$ was constrained to ensure the peak location was at least two $3.5$~MeV $\alpha$ Larmor orbits, $d_{\rm l, \alpha}$, away from $\Psi_{\rm fcfs}$, which is later verified with particle tracing codes to minimize prompt losses of the fusion products; see Section \ref{sec:a_alpha_heating_results}. 

The core magnet coil is expected to produce extreme Lorentz forces in a power plant scale device. The final design is constrained to have a peak von Mises stress of less than $700$~MPa to prevent yielding of the structural materials and to ensure a strain in the REBCO tape below $0.4$\% \cite{gaifullin:2023, shin:2007}. To aid in this, the core magnet coil REBCO tape fill fraction was limited to $<40$~\% to allow for an adequate cross section of steel within the winding pack. The width of the low-field region is then set to be $150$~mm to allow adequate space for the on-board superconducting power supply outlined in Section~\ref{sec:dp_fp} and any other sensitive on-board systems. Finally, the maximum surface temperature of the neutron shield tungsten tiles is limited to be less than $2500$~K to prevent large thermal creep rates.

\begin{table}[b!]
    \centering
    \resizebox{\columnwidth}{!}{
        \begin{tabular}{l l c c l}
        \hline\hline
        Name & Symbol & Reactor A & Reactor B & Units \rule{0pt}{4ex}\\ [1.5ex]
        \hline
        Net electric power & $P_{\rm net}$ & $208$ & $74.5$ & MW \rule{0pt}{2.6ex}\\ [0.5ex]
        Auxiliary heating power & $P_{\rm aux}$ & $44.5$ & $15.8$ & MW \\ [0.5ex]
        Target Q & $Q_{\rm sci}$ & \multicolumn{2}{c}{15} & \\ [0.5ex]
        Core magnet duty cycle & $f_{\rm d}$ & $90.1$ & $90.2$ & \% \\ [0.5ex]
        Core magnet outer radius & $R_{\rm cm}$ & $7.1$ & $6.1$ & m \\ [0.5ex]
        First wall radius & $R_{\rm fw}$ & $20.6$ & $16.9$ & m \\ [0.5ex]
        Outer chamber radius & $R_{\rm ch}$ & $25.9$ & $21.8$ & m \\ [0.5ex]
        Core Magnet Surface Area & $A_{\rm cm}$ & $320$ & $221$ & m$^2$ \\ [0.5ex]
        First Wall Surface Area & $A_{\rm fw}$ & $4,020$ & $2,720$ & m$^2$ \\ [0.5ex]
        Tritium breeding ratio & & \multicolumn{2}{c}{1.1} & \\ [0.5ex]
        Plant availability factor & & \multicolumn{2}{c}{96} & \% \\ [0.5ex]
        \hline
        Component & Material &  &  &  \rule{0pt}{4ex}\\ [1.5ex]
        \hline
        Outer VV & Reinforced Concrete & $38,700$ & $23,400$ & tonnes \rule{0pt}{2.6ex}\\ [0.5ex]
        Tritium Breeding Blanket & Li$_2$O & $3,490$ & $2,340$ & tonnes \\ [0.5ex]
        Inner VV & Inconel 718 & $325$ & $216$ & tonnes \\ [0.5ex]
        Neutron Shield Tiles & Tungsten & $1,760$ & $1,100$ & tonnes \\ [0.5ex]
        ${\rm B}_4{\rm C}$ Shield & ${\rm B}_4{\rm C}$ & $82.3$ & $51.9$ & tonnes \\ [0.5ex]
        WC Shield & WC & $168$ & $99.1$ & tonnes \\ [0.5ex]
        Core Magnet Structure & SS316LN & $351$ & $162$ & tonnes \\ [0.5ex]
        Coil Conduit & SS316LN and Copper & $199$ & $76.9$ & tonnes \\ [0.5ex]
        REBCO Tape & REBCO & $4,320$ & $2,550$ & km \\ [0.5ex]
        \hline
        \textbf{CM total} &  & $2,560$ & $1,490$ & tonnes \rule{0pt}{2.6ex}\\ [0.5ex]
        \textbf{Reactor total} &  & $45,100$ & $27,400$ & tonnes \rule{0pt}{2.6ex}\\ [0.5ex]
        \hline\hline
    \end{tabular}
    }
    \caption{High level reactor target parameters and material usage for two optimized power plant designs with constraints placed on the overnight capital cost and LCOE (Section \ref{sec:methods}).}
    \label{tab:ReactorOverview}
\end{table}

\section{Design Points and Analysis}\label{sec:analysis}

\begin{figure}[t!]
    \centering
    \includegraphics[width=1\linewidth]{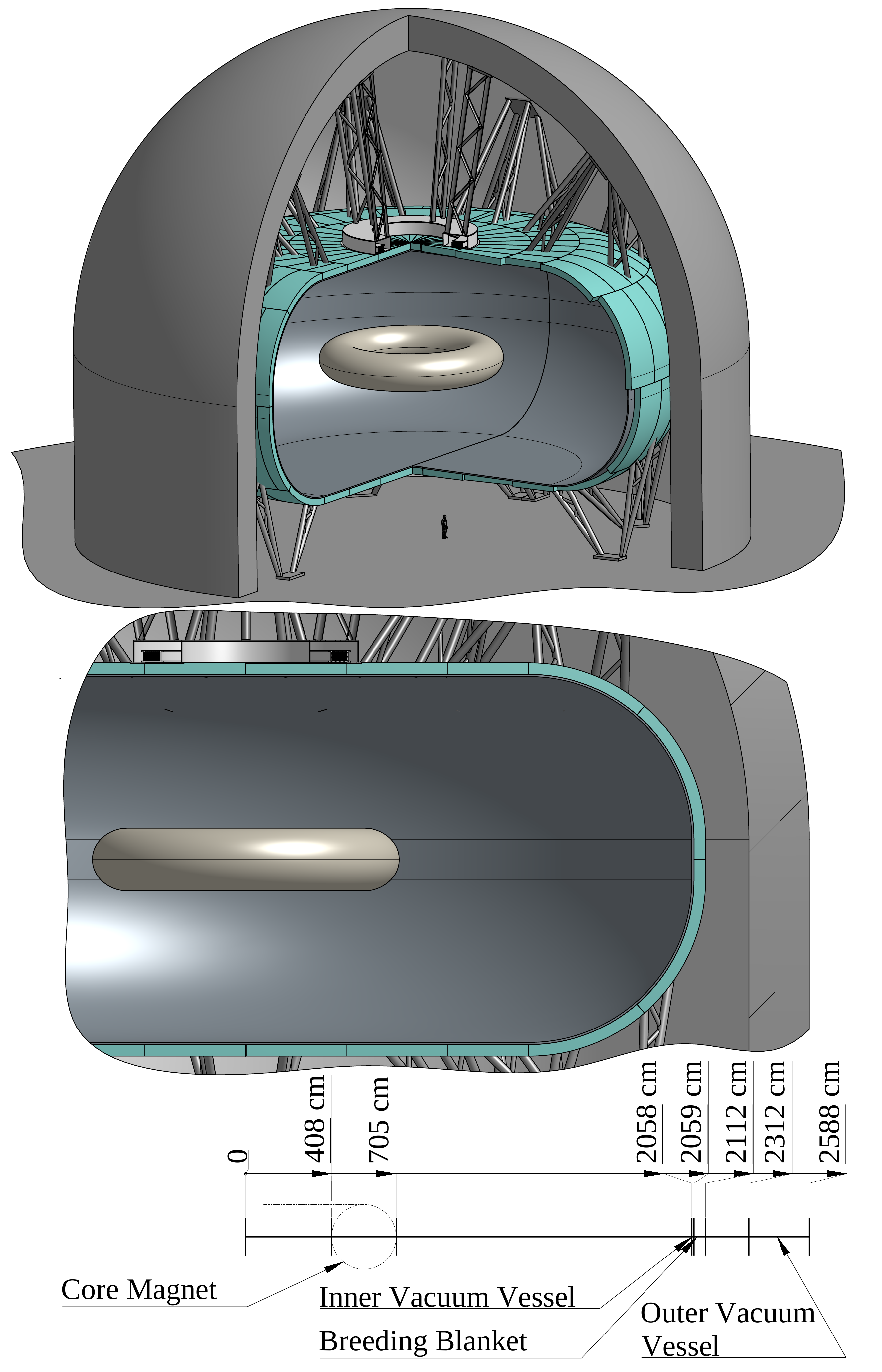}
    \caption{Cross section and radial build of Reactor A (208 MWe).}
    \label{fig:ReactorCrossSection}
\end{figure}

The optimization process produces significantly different designs for the two overnight capital cost constraints described in Section~\ref{sec:m_performance_constraints}. An overview of the two design points is given in Table~\ref{tab:ReactorOverview}. Reactor A is a $208$~MWe plant with an outer core magnet radius, defined to include neutron shielding, of $7.1$~m, a first wall/limiter radius of $20.6$~m, and an outer vacuum vessel radius of $25.9$~m. This is a similar output power to ITER \cite{aymar:2002} and Commonwealth Fusion System's 2016 ARC \cite{sorbom:2015}, making it a useful point of comparison. The overall size of Reactor A is much larger than either of the aforementioned tokamaks, however the majority of this space, as shown in Fig.~\ref{fig:ReactorCrossSection}, is comprised of the extremely simple vacuum vessel. The core magnet, which is the most complex and expensive part of the reactor, is the same physical scale as the magnets that comprise the ARC tokamak \cite{sorbom:2015} (Fig.~ \ref{fig:magnet_scale}). Therefore the capital cost of the plant remains competitive. Reactor B, on the other hand, generates a lower power of $75$~MWe with a smaller $6.1$~m radius core magnet and a $21.8$~m radius vacuum vessel. At this output power Reactor B is more suited for industrial applications instead of standalone grid power generation. The lower overnight capital cost may make Reactor B the more appealing choice as a FOAK fusion power plant.

\begin{figure}[t!]
    \centering
    \includegraphics[width=1\linewidth]{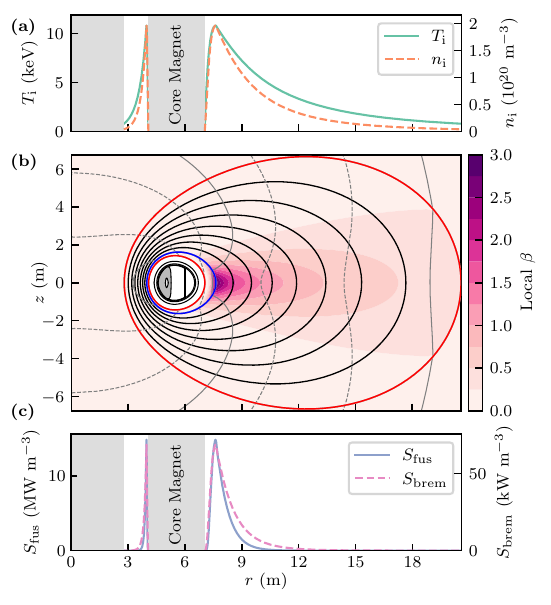}
    \caption{Plasma equilibrium of the Reactor A ($208$~MWe) design point. (a) The ion temperature and density profiles along the midplane. (b) The calculated Grad-Shafranov equilibrium showing contours of magnetic flux. Contours of magnetic field strength have also been included as the dash-dotted lines to highlight the effects of $\beta_0$. (c) Fusion and bremsstrahlung power density profiles along the midplane.}
    \label{fig:DARC_Equilibrium}
\end{figure}

The key plasma parameters of the target equilibrium for both reactors are presented in Table~\ref{tab:Plasma Parameters}. Both reactors operate with a core temperature of $\sim10$~keV (see Section~\ref{sec:a_scaling_results}) and require an edge temperature pedestal of $790$~eV. As shown in Fig.~\ref{fig:DARC_Equilibrium}, the total fusion power is significantly higher on the outboard side of the core magnet than in the bore. This difference is a result of the large flux expansion that is characteristic of dipole plasmas. As temperature and density are both purely functions of the poloidal magnetic flux (see Section~\ref{sec:dp_stability}), this expansion leads to larger volumes of fusion reactions on the low field side of the core magnet. As a result, the total fusion power on the outboard side is $3-4$ times higher than in the magnet bore. 

Levitated dipoles are often quoted as being high $\beta$ devices \cite{hasegawa:1990, kesner:2003}, however due to the large inhomogeneity in the plasma pressure profile this is only true for a local definition of $\beta$. Both reactors display $\beta_0=2.9$ which, as described in Section~\ref{sec:dp_large_beta}, is in the optimal fusion power range for the given magnet configuration. On the other hand, $\langle\beta\rangle$ as defined in Eq.~\eqref{eq:beta_p} is much lower, sitting at just $\langle\beta\rangle=4.37$~\% for Reactor A and $\langle\beta\rangle=4.84$~\% for Reactor B. These values are coincidentally typical of an ARC class tokamak \cite{sorbom:2015}. Unlike a tokamak, driving the plasma to higher $\langle\beta\rangle$ results in a loss of confinement through infinite local $\beta_0$ and the resultant plasma expansion (Section~\ref{sec:dp_large_beta}) instead of the excitement of MHD instabilities. 

\begin{table}[b!]
    \centering
    \resizebox{\columnwidth}{!}{
    \begin{tabular}{l l c c l}
        \hline\hline
        Name & Symbol & Reactor A & Reactor B & Units \rule{0pt}{4ex}\\ [1.5ex]
        \hline
        Fusion power & $P_{\rm fus}$ & $667$ & $237$ & MW \rule{0pt}{2.6ex}\\ [0.5ex]
        Peak plasma pressure & $p_0$ & $0.68$ & $0.52$ & MPa \\ [0.5ex]
        Peak ion temperature & $T_{\rm i,0}$ & $10.9$ & $9.77$ & keV \\ [0.5ex]
        Peak electron density & $n_{\rm i,0}$ & $1.95\times10^{20}$ & $1.65\times10^{20}$ & m$^{-3}$ \\ [0.5ex]
        Peak local $\beta$ & $\beta_0$ & \multicolumn{2}{c}{$2.9$} & \\ [0.5ex]
        Peak pressure radius & $R_0$ & $7.6$ & $6.6$ & m \\ [0.5ex] 
        Peak pressure flux & $\psi_0$ & $0.11$ & $0.141$ & \\ [0.5ex]
        $B$ field at $R_0$ & $B_0$ & $0.77$ & $0.68$ & T \\ [0.5ex]
        Edge ion temperature & $T_{\rm i,lcfs}$ & \multicolumn{2}{c}{$790$} & eV \\ [0.5ex]
        Edge electron density & $n_{\rm i,lcfs}$ & \multicolumn{2}{c}{$3.91\times10^{18}$} & m$^{-3}$ \\ [0.5ex]
        $14.1$ MeV neutron power & $P_{\rm n}^{\rm DT}$ & $532$ & $189$ & MW \\ [0.5ex]
        $2.45$ MeV neutron power & $P_{\rm n}^{\rm DD}$ & $0.51$ & $0.18$ & MW \\ [0.5ex]
        Bremsstrahlung power & $P_{\rm brem}$ & $21.5$ & $9.5$ & MW \\ [0.5ex]
        Plasma stored energy & $U_{\rm p}$ & $543$ & $263$ & MJ \\ [0.5ex]
        Global $\beta$ & $\langle\beta\rangle$ & $4.37$ & $4.84$ & \% \\ [0.5ex]
        Plasma Volume & $V_{\rm p}$ & $1.36\times10^4$ & $7.41\times10^3$ & m$^3$ \\ [0.5ex]
        Energy confinement time & $\tau_{\rm e}$ & $3.5$ & $5.9$ & s \\ [0.5ex]
        \hline\hline
    \end{tabular}
    }
    \caption{Reactor plasma equilibrium parameters for the Reactor A ($208$~MWe) and Reactor B ($74$~MWe) design point plants.}
    \label{tab:Plasma Parameters}
\end{table}

The total stored energy in Reactor A is $U_{\rm p} = 543$~MJ. From this the energy confinement time required in this device was calculated using Eq.~\eqref{eq:tau_e_estimate} to be $\tau_{\rm e}=3.5$ s. This is lower than the energy confinement time required for Reactor B of $5.9$ s. This implies that Reactor A is a more conservative design of a $Q_{\rm sci}=15$ device, which is discussed further in Section~\ref{sec:a_scaling_results}. For now it is worth noting the implications of such a high energy confinement time. Due to the convective cells that form in the bad curvature region the particle confinement time in a dipole plasma will be short. In ideal conditions, marginal stability to interchange modes will result in these convective cells transporting no energy. However, due to losses at the plasma edge there must be some energy transport to maintain the MHD stable profile. It is expected that the resultant energy confinement time, which we have presented here, will be an order of magnitude larger than the particle confinement time \cite{kesner:2003}. This will allow for efficient removal of ash and refueling to occur near $\Psi_{\rm lcfs}$ without excess loss of plasma energy. 

The extra size and simplicity of the reactor vacuum vessel offers additional benefits when it comes to plant access and maintainability (Section~\ref{sec:dp_RAMI}). The key advantage here is that the core magnet is completely decoupled from the vacuum chamber, meaning it can be removed and replaced without the disassembly of the whole plant. Although they are not the focus of this study, this modularity also applies to other systems in the reactor such as the breeder blanket, top magnet coil, and limiter. The blanket in particular is mounted to the outside of the inner vacuum vessel which allows easy access for maintenance and replacement. This replacement could be done in stages throughout the life of the plant with minimal impact on the plant up time, which is important as it allows the use of solid blanket materials. The large outer vacuum vessel is constructed using reinforced concrete at similar sizes to previously constructed vacuum vessels \cite{sorge:2013}.

In order to focus on the key engineering challenges in a levitated dipole fusion power plant, the details of the limiter and top magnet have not been considered. The strength and stored energy of the core magnet mean that it will only require a small secondary field provided by the top magnet in order to generate sufficient levitation force. This immediately ensures the top magnet will be a small fraction of the core magnet cost and therefore will not affect the cost of the overall plant significantly. As mentioned in Section~\ref{sec:dp_edge}, we also expect that a fully realized levitated dipole reactor will have a diverted plasma in order to achieve the edge conditions presented in this study, hence presenting a design for a limiter would be redundant.

\begin{table}[b!]
    \centering
    \resizebox{\columnwidth}{!}{
    \begin{tabular}{l l c c l}
        \hline\hline
        Name & Symbol & Reactor A & Reactor B & Units  \rule{0pt}{4ex}\\ [1.5ex]
        \hline
        Coil outer radius & $R_{\rm c}$ & $5.3$ & $4.7$ & m \rule{0pt}{2.6ex}\\ [0.5ex]
        Peak field on conductor & $B_\text{max}$ & $23.0$ & $21.8$ & T\\ [0.5ex]
        Current density & $J_\text{op}$ & $65.1$ & $101$ & A/mm$^2$ \\ [0.5ex]
        Inductance & $L$ & $48.1$ & $21.8$ & H \\ [0.5ex]
        Total current & $I_\text{tot}$ & $52.8$ & $35.7$ & MA-turns \\ [0.5ex]
        Stored energy & $U_\text{mag}$ & $20.8$ & $9.47$ & GJ \\ [0.5ex]
        Terminal current & $I_\text{op}$ & $29.4$ & $29.5$ & kA \\ [0.5ex]
        Peak structure Von Mises stress & $\sigma_{\rm vm}$ & $670$ & $663$ & MPa \\ [0.5ex]
        Average structure Von Mises stress & $\langle\sigma_{\rm vm}\rangle$ & $543$ & $542$ & MPa \\ [0.5ex]
        Peak magnet axial stress & $\sigma_{zz}$ & $-547$ & $-482$ & MPa \\ [0.5ex]
        Peak magnet hoop stress & $\sigma_{\varphi\varphi}$ & $477$ & $459$ & MPa \\ [0.5ex]
        Peak magnet tensile strain & $\varepsilon_{\rm t}$ & $0.35$ & $0.33$ & \% \\ [0.5ex]
        Operating temperature & $T_\text{op}$ & \multicolumn{2}{c}{$30$} & K \\ [0.5ex]
        Coolant & & \multicolumn{2}{c}{Neon slush} & \\ [0.5ex]
        Float time & $t_\text{float}$ & $45.5$ & $46.1$ & min \\ [0.5ex]
        \hline\hline
    \end{tabular}
    }
    \caption{Core magnet design parameters for the Reactor A ($208$~MWe)
and Reactor B ($75$~MWe) design point plants.}
    \label{tab:MagnetParameters}
\end{table}

\begin{figure}[b!]
    \centering
    \includegraphics[width=1\linewidth]{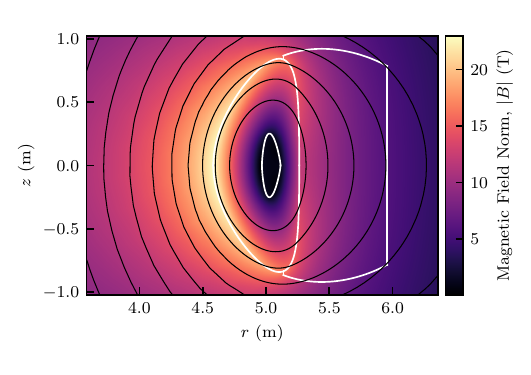}
    \caption{The magnetic field produced by the Reactor A ($208$~MWe) core magnet coil. The coil produces a peak field of $23.0$~T at the inboard midplane while simultaneously creating a low-field region in its core with field strengths $<100$~mT. The shape of the magnet was optimized in part to minimize induced stresses.}
    \label{fig:darc_magnetic_field}
\end{figure}

Also presented in Table~\ref{tab:ReactorOverview} is a summary of the total masses of each component in the reactor. In Reactor A, the core magnet assembly is predicted to weigh in excess of $2500$~tonnes, with the majority of the mass comprised of the neutron shield tungsten tiles. This mass does not pose much of a concern for the levitation of the core magnet, however, future designs of levitated dipole fusion power plants should aim to minimize this tungsten use. The sensitivity of the overall reactor size to the thickness of the neutron shield would allow even modest increases in material neutron attenuation performance to result in significant savings in overall shield mass. Therefore, the material selection for the core magnet neutron shield is still an area of active research.

\subsection{Magnet Design}\label{sec:a_magnet_results}

\begin{figure}[b!]
    \centering
    \includegraphics[width=1\linewidth]{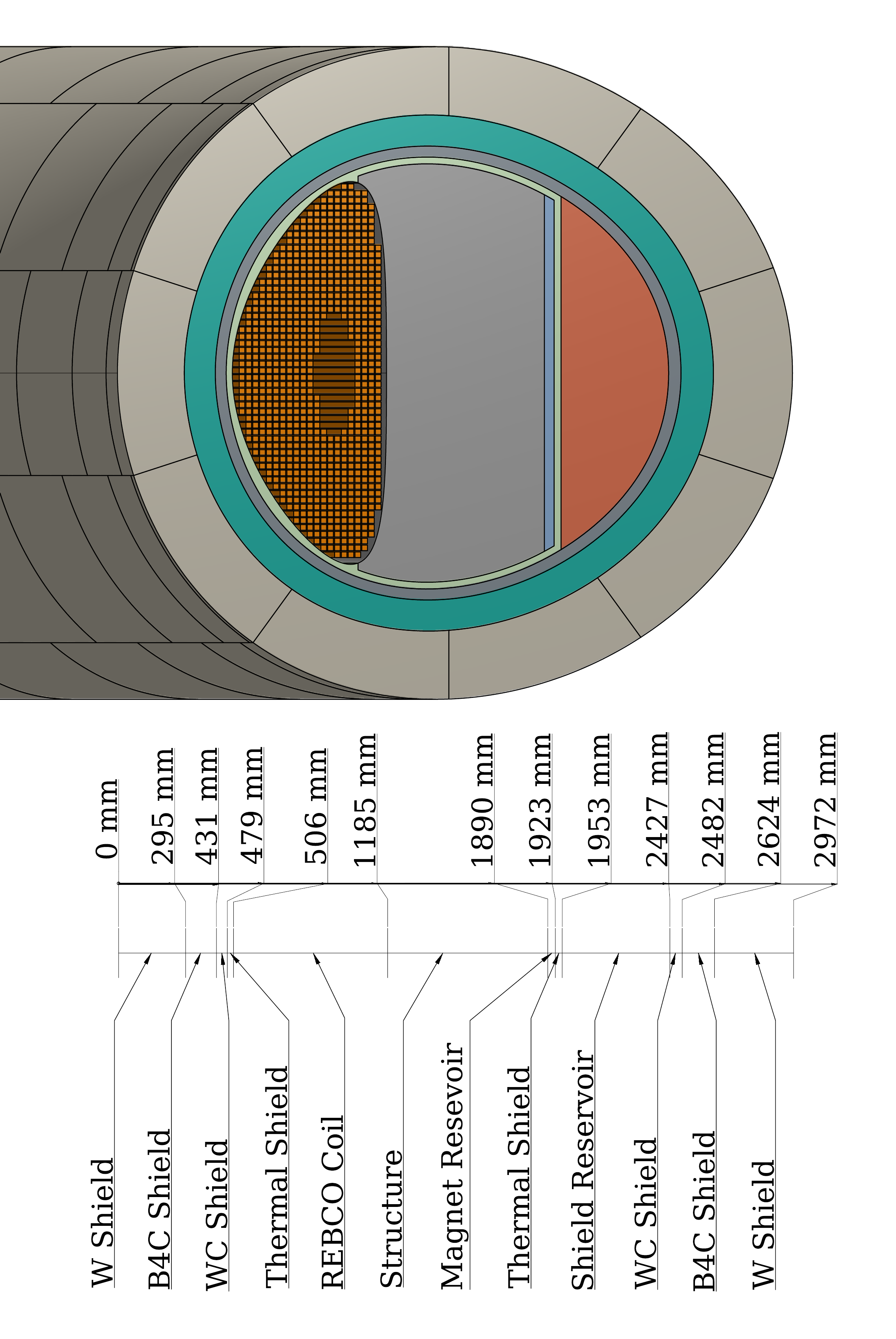}
    \caption{Cross section of the Reactor A ($208$~MWe) core magnet and radial build. A thermal break between the outer tungsten shield and the ${\rm B}_4{\rm C}$ layer is assumed, but is not shown here. The thickness of the shield required to get acceptable core magnet coil lifetimes results in only a small amount of neutron heating. Therefore the required volume of cryogen slush to achieve adequate float times is small.}
    \label{fig:DARC_cross_section}
\end{figure}

\begin{figure*}[!t]
    \centering
    \includegraphics[width=1\linewidth]{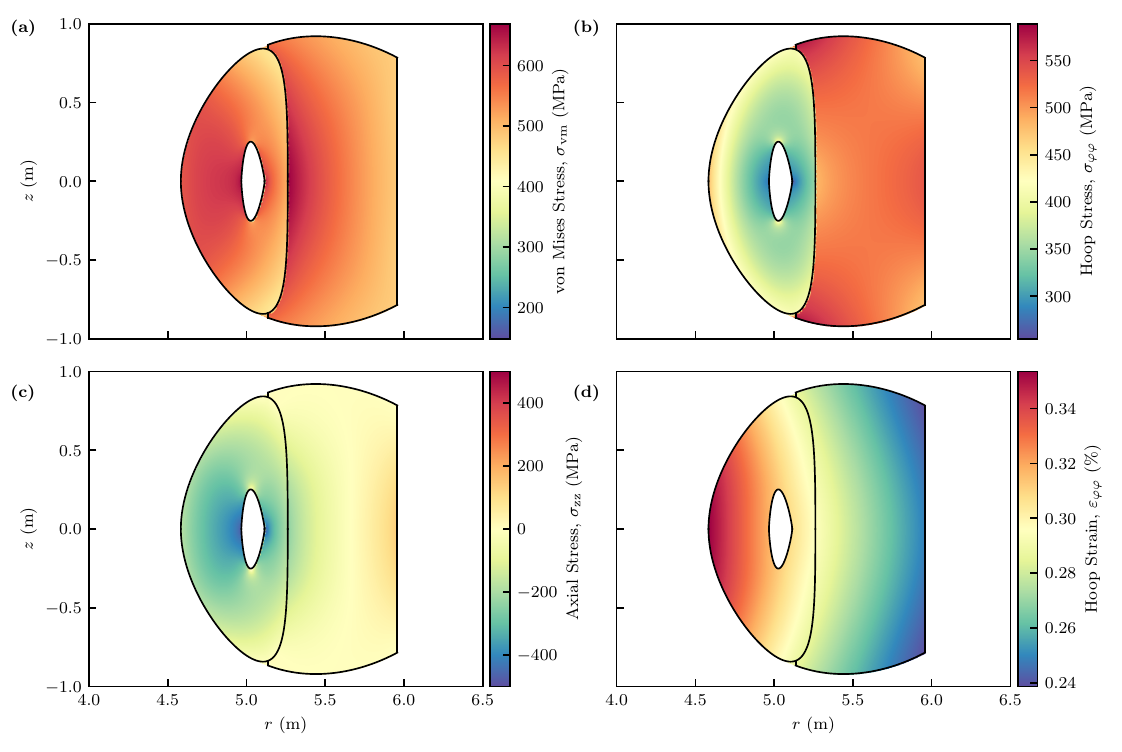}
    \caption{Calculated (a) Von Mises stress, (b) hoop stress, (c) axial stress, and (d) hoop strain for a 2D axisymmetric finite element linear elastic model of Reactor A ($208$~MWe). The low-field region at the core of the core magnet coil is assumed to be able to carry load only in the $rz$ plane, however it has been excluded from these plots to avoid confusion.}
\label{fig:darc_stress_strain}  
\end{figure*}

The core magnet required to confine the plasma is comprised of a single large REBCO coil whose cross section has been optimized to offer the best plasma performance constrained by material yield stress limits. The design parameters for the core magnets of both reactors are presented in Table~\ref{tab:MagnetParameters}. For Reactor A, the core magnet coil generates a peak field of $23.0$~T and has a total stored energy of $20.8$~GJ, similar to that of the full ARC toroidal field magnet system \cite{hartwig:2024}. As shown in Fig.~\ref{fig:DARC_cross_section} this coil is mounted to a substantial 316LN stainless steel support with space for a reservoir for the slush cryogen coolant. This assembly is placed within a thin cryostat to thermally shield it from the $\sim 600$ $^\circ {\rm C}$ inner surface of the neutron shield.

The REBCO coil is constructed using a cable-in-conduit (CICC) style cable where a stack of 6 mm wide REBCO tape is soldered to a copper channel and then wrapped in a square cross section steel jacket. The turns of the coil are then assumed to be welded together with insulation only placed between adjacent winding layers, leaving adjacent turns to be non-insulated. This configuration allows for both higher coil stiffness for more effectively transferring stress to the structural over-band, and better cooling and resilience in the event of a quench. The exact proportions of steel to copper and the specifics of the cooling method are out of the scope of this study. About $20 $\% of the coil is designated as sacrificial, meaning that it will be replaced at a regular cadence due to damage from the fusion neutrons. This section will be mechanically separate from the remainder of the coil to ensure it can be removed easily. 

The REBCO cable will be designed to carry $\sim30$~kA of current continuously while in operation. This current is supplied by a superconducting power supply mounted in the low-field region in the center of the coil. This power supply includes components that will not function in the high field environment generated by the core magnet coil. The shape of the low-field region is optimized to reduce the field to a level $<100$~mT where passive shielding methods become possible. Fig.~\ref{fig:darc_magnetic_field} shows that by choosing the correct coil and low-field region shape, the coil itself acts to create a region with field strengths two orders of magnitude lower than the peak coil field. The inclusion of the superconducting power supply allows the magnet to stay charged while levitating and is a key component in reducing the time the core magnet spends docked. One key limiting factor of superconducting power supplies is their inability to produce high voltages. At the time of writing, the highest performance flux pumps produce voltages ranging in the millivolts \cite{Geng:2025,Geng:2025}, well below what is needed to charge a magnet of the sizes described here in a practical amount of time. For this reason, the superconducting power supply is only used to maintain the current in the coil, in a quasi-persistent state, while external semi-conducting power supplies are used in the rare occasions this magnet will need to be charged. 

The coil will operate at a temperature of $30$~K set by the melting point of the chosen cryogen with some margin, in this case neon with a melting point of $24.6$~K \cite{ekin:2006} which was chosen for its superior latent heat capacity. The neon will be stored on the magnet in the form of a solid-liquid slush mixture which will transition to be fully liquid as it absorbs energy. The volume of the reservoir determined through the optimization process is sufficient to allow for $>45$~minutes of levitation before the core magnet will need to be docked and the melted slush pumped out and replaced. The overall plant duty cycle for both reactors is high, $f_{\rm d} > 90$~\%, which offers the best tradeoff between increasing the float time and reducing the volume available for the core magnet coil and structural over-band. The amount of neon needed to achieve this levitation time is small, which is a result of the lifetime of the REBCO conductor in the core magnet coil being the limiting factor when determining the thickness of the neutron shield (Section~\ref{sec:a_neutron_shielding_results}). The remaining heat loads from the attenuated neutrons and photons, electrical heating, and conduction from the neutron shield only sum to a few tens of kilowatts as shown later in Table~\ref{tab:EnergyBalance}. Another possible cryogen would be hydrogen which is significantly cheaper than neon and has a lower melting point of $14.0$~K \cite{ekin:2006}, but needs $\sim 5$ times the volume to store the same amount of energy. This increased volume would have a small impact on the allowed volume for structural support and hence would result in slightly larger reactors. However if procuring and maintaining a supply of neon proves challenging it would be a viable alternative.

The expected stresses and strains in a homogenized version of the core magnet were modeled in COMSOL to verify that the conductor was not pushed beyond its mechanical strain limit. The coil material was treated as an arbitrary mixture of copper and steel with a Young's modulus of $150$~GPa. We have also assumed that the low-field region will be able to carry stress in the $rz$ plane, but not in the $\varphi$ direction, to aid in the overall stress distribution. The structural over-band was assumed to be solid stainless 316LN for the purposes of this calculation with the expectation that some of the material will be removed to reduce weight and allow space for magnet services. 

The results of this modeling, presented for Reactor A in Fig.~\ref{fig:darc_stress_strain}, show the peak hoop strain in the conductor is $0.35$~\% for Reactor A and $0.33$~\% for Reactor B. These are both below the mechanical limit of $0.4$~\% \cite{gaifullin:2023} valid for most REBCO tapes. The peak von Mises stress in both reactors also remained below $75$~\% of the cryogenic yield stress of 316LN of $1050$~MPa \cite{nyilas:2004}. The average von Mises stress in Reactor A is $543$~MPa and in Reactor B is $542$~MPa, which is low enough to allow for significant material to be removed without compromising structural integrity. This would reduce the overall mass of the magnet, but would mainly serve to allow for cooling channels and docking infrastructure. One of the main concerns with previous proposed levitated dipole fusion reactors was the need of novel structural materials to generate reasonable output powers \cite{kesner:2003}. However, here we have shown a magnet design that can be built with contemporary materials and traditional manufacturing methods, substantially reducing cost and technology risk. 

\subsection{Fast Ion Confinement}\label{sec:a_alpha_heating_results}

\begin{figure*}[t!]
    \centering
    \includegraphics[width=1\linewidth]{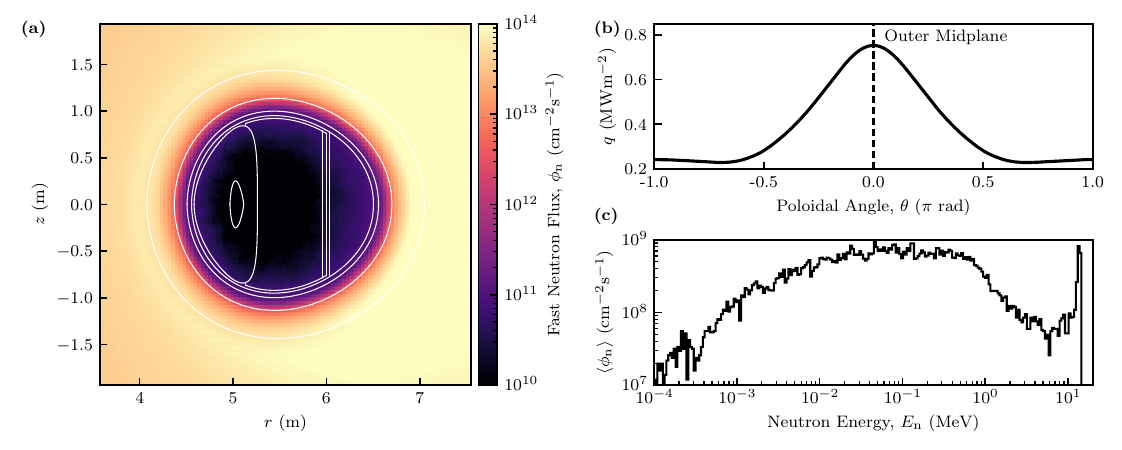}
    \caption{Results from OpenMC calculations displaying (a) the fast ($E_{\rm n}>0.1$~MeV) neutron flux, (b) the neutron and radiation wall loading on the neutron shield, and (c) the average neutron flux energy spectrum for Reactor A ($208$~MWe). The neutron shield sees a peak wall loading of $0.753$~MW\,m$^{-2}$ and results in a peak neutron flux on the core magnet coil of $8.05\times10^{10}$~cm$^{-2}$s$^{-1}$ which, due to the way in which the neutron shield was designed, is also the average flux at the inner surface of the neutron shield. Neutron moderation in the neutron shield results in the majority of the remaining neutron power being in the $0.01$-$1$~MeV range.}
    \label{fig:neutron_flux}
\end{figure*}

Prompt $\alpha$-particle losses were calculated using ASCOT5, a test-particle orbit-following code \cite{varje:2019}. ASCOT5 simulations are based on a volume preserving algorithm integrating particle orbits through a fixed time step, which was set to $1~\rm{ns}$, or $\sim5\%$ of the shortest gyro-period in both reactors. Coulomb collisions with the deuterium and tritium ions in the background plasma are included in the simulations, allowing fast $\alpha$-particles to thermalize, with a cutoff set at twice the ion temperature. In each simulation $12,288$ test-particles were sampled from an energy-pitch distribution generated from the reaction rate of Maxwellian reactant pairs \cite{siren:2017}. 

Losses are particles which pass through either $\Psi_{\rm fcfs}$ or $\Psi_{\rm lcfs}$, and are defined as prompt-loss if they still have at least $80~\%$ of their initial energy when lost. The ratios of confined $\alpha$-energy to total $\alpha$-energy in reactors A and B are $>99$~\%, and the corresponding values for the energy lost are given in Table \ref{tab:NeutronShieldParamters}. The peak wall loads are $<1$~kW\,m$^{-2}$, which are two  orders of magnitude less than the neutron loading. Therefore, neutron heating is the primary constraint when designing the shield, as discussed in Section~\ref{sec:a_neutron_shielding_results}. In total Reactor A sees $106$~kW of prompt $\alpha$ heating on the neutron shield, and Reactor B sees $58$~kW.

Steep profiles are a key signature of dipole equilibria, where peak pressure, density and temperature values in the core are much greater than in the bulk plasma. $\alpha$ heating on either side of the peak location, which separates the good-curvature and bad-curvature regions, has a significant effect on the power balance in each region. Heating in the good-curvature region must be balanced by losses to preserve steady state. This is an ongoing area of active research and will be discussed in future works.

\subsection{Neutron Transport}\label{sec:a_neutron_shielding_results}

\begin{figure}[b!]
    \centering
    \includegraphics[width=1\linewidth]{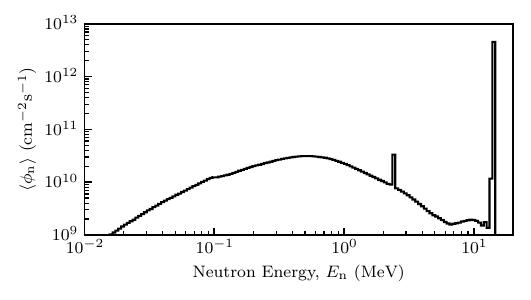}
    \caption{Average neutron flux energy spectrum at the inner surface of the first wall of Reactor A ($208$~MWe). Significant neutron scattering and endothermic multiplication in the tungsten neutron shield results in a population of neutrons with energies in the order of $1$~MeV and similar fluxes to the $2.5$~MeV DD neutrons.}
    \label{fig:fw_neutron_spectrum}
\end{figure}

\begin{figure*}[t!]
    \centering
    \includegraphics[width=1\linewidth]{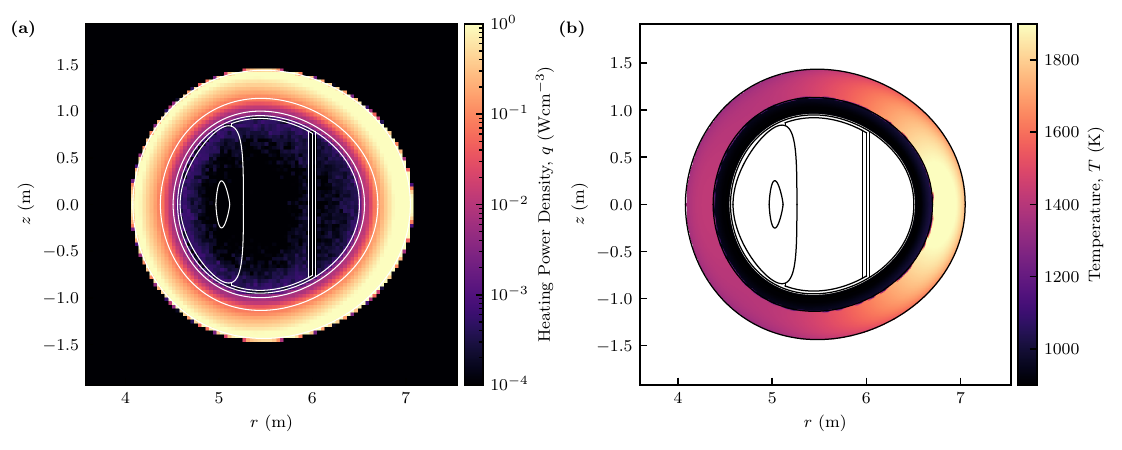}
    \caption{(a) OpenMC calculations of neutron and photon heating power density in the Reactor A ($208$~MWe) core magnet. $92.2$~\% of the heating is localized in the outer neutron tiles. (b) Modeled temperature of the neutron shield including bremsstrahlung and prompt loss heating. A thermal break was modeled between the outer tungsten tiles and the middle ${\rm B}_4{\rm C}$ layer to encourage radiative cooling. The maximum steady state temperature in this scenario is $1950$~K.}
    \label{fig:neutron_heating}
\end{figure*}

The OpenMC code \cite{romano:2015} was used to to investigate the effectiveness of the neutron shielding mounted on the core magnet. The key results from this simulation are presented in Table~\ref{tab:NeutronShieldParamters}. A simplified version of the CAD model shown in Fig.~\ref{fig:DARC_cross_section} was passed through the DAGMC \cite{wilson:2010} workflow for meshing and material assignment. The ENDF/BVII.1 \cite{chadwick:2011} material library was used to calculate the material reaction cross sections. Unfortunately this database does not include an entry for neon hence the cryogen volume was treated as a vacuum. This should not affect the results substantially as the number density of the neon slush is low and it also only comprises a small portion of the physical cross section of the core magnet. The core magnet coil was modeled as a $16.5~\%$ - $16.5~\%$ - $16.5~\%$ - $50.5~\%$ mix of YBCO superconducting tape, PbSn solder, Copper, and 316LN stainless steel. The REBCO tape was also assigned the appropriate mix of YBCO, copper, and Hastelloy commonly found in tape from the larger REBCO tape suppliers \cite{molodyk:2021}. The structural over-band was modeled assuming $80$~\% density to account for cooling channels and docking infrastructure. Finally, the neutron source was modeled as a fine grid of isotropic mono-energetic ring sources spanning the plasma region depicted in Fig.~\ref{fig:DARC_Equilibrium} with the $2.45$~MeV DD and $14.1$~MeV DT neutrons treated as separate sources. The source weighting was then calculated using the plasma equilibrium profile to get the fully resolved neutron source distribution.

Some important results from this model are given in Table~\ref{tab:NeutronShieldParamters}. The total heating in the cryogenic region from both neutrons and secondary photons is $14.1$~kW for Reactor A, and $7.67$~kW for Reactor B. In both cases the heating power is a sufficiently small percentage of the total fusion power required for the plant to be power positive. The factor requiring such a thick shield is therefore the lifetime of the sacrificial portion of the core magnet coil. For Reactor A, a $475$~mm thickness shield at the point closest to the core magnet is sufficient to reduce the fast neutron flux by four orders of magnitude to $8.05\times10^{10}$~cm$^{-2}$s$^{-1}$. The neutron flux spatial distribution and attenuated neutron flux energy spectrum at the inner surface of the shield is given in Fig.~\ref{fig:neutron_flux}. The majority of the neutron energy is moderated down to a band between $0.01$ and $1$~MeV, which is a similar spectra to the neutron sources used to test the lifetime of REBCO samples \cite{fischer:2018}. At these energies the critical current of REBCO tape drops by $5$~\% after a fluence of $3\times10^{18}$~cm$^{-2}$ \cite{fischer:2018}.

\begin{table}[b!]
    \centering
    \resizebox{\columnwidth}{!}{
    \begin{tabular}{l l c c l}
        \hline\hline
        Name & Symbol & Reactor A & Reactor B & Units \rule{0pt}{4ex}\\ [1.5ex]
        \hline
        Peak Neutron Flux & $\phi_{\rm n}$ & $1.78\times10^{14}$ & $8.60\times10^{13}$ & cm$^{-2}$s$^{-1}$\rule{0pt}{2.6ex}\\ [0.5ex]
        First Wall Neutron Flux & $\langle \phi_{\rm n} \rangle_{\rm fw}$ & $2.69\times10^{12}$ & $2.43\times10^{12}$ & cm$^{-2}$s$^{-1}$\\ [0.5ex]
        Neutron Flux on REBCO Coil & $\langle \phi_{\rm n} \rangle_{\rm cm}$ & $8.05\times10^{10}$ & $6.97\times10^{10}$ & cm$^{-2}$s$^{-1}$\\ [0.5ex]
        Max Neutron Shield Wall Loading & $q_{\rm ns}$ & $0.753$ & $0.350$ & MWm$^{-2}$ \\ [0.5ex]
        Max First Wall Loading & $q_{\rm fw}$ & $0.198$ & $0.097$ & MWm$^{-2}$ \\ [0.5ex]
        Blanket Neutron Energy Fraction & $f_{\rm b}$ & $80.6$ & $81.7$ & \% \\ [0.5ex]
        Total Neutron Shield Heating & $P_{\rm ns}$ & $90$ & $30$ & MW \\ [0.5ex]
        Total Cryogenic Heating & $P_{\rm cryo}$ & $14.1$ & $7.67$ & kW \\ [0.5ex]
        $\alpha$ Prompt Losses & $P_{\rm prompt}$ & $106$ & $58$ & kW \\ [0.5ex]
        Maximum Shield Temperature & $T_{\rm ns}$ & $1950$ & $1540$ & K \\ [0.5ex]
        Neutron Shield Thickness & $d_{\rm ns}$ & $475$ & $449$ & mm\\ [0.5ex]
        Sacrificial REBCO Lifetime & & $1.2$ & $1.4$ & years\\ [0.5ex]
        Core Magnet Coil Lifetime & & $12$ & $14$ & years\\ [0.5ex]
        \hline\hline
    \end{tabular}
    }
    \caption{Reactor neutron loading parameters and neutron shield design parameters for the Reactor A ($208$~MWe) and Reactor B ($75$~MWe) design point plants.}
    \label{tab:NeutronShieldParamters}
\end{table}

With the observed neutron flux magnitudes, the sacrificial section of the core magnet coil will see a $5$~\% reduction in its critical current after just over $1$~year at which point it will need to be replaced. The remaining portion of the REBCO in the core magnet coil will have a lifetime of at least $10$~years. The model used in Fig.~\ref{fig:neutron_flux} assumes the material composition is the same through out the core magnet coil and as a result shows that $\sim50$~\% of the coil would see a high neutron flux. The final coil design will utilize neutron shielding materials in these regions such as tungsten borides or metal hydrides in order to reduce the neutron mean free path length and reduce the high flux area to $20$~\% of the coil cross section. The lifetime of the steel in the structural over-band is expected to be significantly longer than the REBCO and therefore will not be replaced during the operational lifetime of the core magnet. This level of degradation is considered acceptable due to the modularity of the levitated dipole concept. As discussed in Section~\ref{sec:dp_RAMI}, there will be a short two week downtime period each year for maintenance and repair, which is comparable to many other mature power generation methods \cite{EIA:2025}. The core magnet can be removed and replaced with a fresh magnet within this time period, allowing maintenance to take place on the damaged coil external to the reaction chamber, significantly reducing cost and maintaining a high plant availability. The cost of these replacements and the effects of the two week downtime have been included in the pricing model used to constrain the optimization process and were found to not make a significant impact on the economic viability of the plant. 

In total $23.5$~\% of the fusion neutrons pass into the region bounded by the first closed flux surface, $\Psi_{\rm fcfs}$. The majority of the energy carried by these neutrons gets deposited as heat in the shield, however some of the incident energy gets absorbed in endothermic neutron multiplication reactions and a larger portion gets scattered or reflected back out towards the first wall beyond the last closed flux surface, $\Psi_{\rm lcfs}$. The neutron multiplication and scattering in the outer tungsten tiles is sufficient for the total number of neutrons passing through $\Psi_{\rm lcfs}$ to equal $1.05$~times the rate of neutrons being produced in the fusion reaction. The energy of these multiplied and scattered neutrons has been moderated down to the range of $\sim 1$~MeV as shown in Fig.~\ref{fig:fw_neutron_spectrum}. This lower energy will have a small impact on the effectiveness of the tritium breeding blanket as discussed in Section~\ref{sec:dp_tritium_breeding}.

Figure \ref{fig:neutron_flux} shows that the wall loading for Reactor A peaks at a value of $0.753$~MW\,m$^{-2}$ on the outboard side of the of the core magnet. This is significantly lower than equivalent power tokamaks which range from $1-2.5$~MW\,m$^{-2}$ depending on the size of the device \cite{sorbom:2015, aymar:2002}. This results in the tungsten outer layer of the neutron shield reaching the $1$~MW-year\,m$^{-2}$ limit \cite{NASEM:2021} after $1.3$~years on the outboard side and $4.4$~years on the inboard side. However, unlike those devices, all the heat deposited in the shield cannot be actively extracted until the magnet is docked. In total, Reactor A experiences $90$~MW of neutron induced heating and $8$~MW of radiative heating from the plasma. The heating from isotope decay has not been included in this model as it is only expected to contribute $\sim 1$~W\,m$^{-3}$ after long periods of neutron irradiation \cite{windsor:2022}. 

The neutron heating, as shown in Fig.~\ref{fig:neutron_heating}, penetrates further into the shield. During a levitation cycle, the majority of the energy deposited in the shield is radiated away at the surface to the assumed $600~^\circ {\rm C}$ first wall. Practicalities such as tritium retention may change this temperature in future designs. A 2D heat transfer model was constructed in COMSOL to calculate the required shield temperatures to support this radiative cooling. The neutron heating was exported from OpenMC and applied as a volumetric source, while the bremsstrahlung and prompt $\alpha$ heating were applied as boundary heat load. The outer surface of the shield was assumed to have an emissivity of $1$ and a thermal break with a $10$~W\,m$^{-2}$K$^{-1}$ conductance was inserted between the outer tungsten tiles and the ${\rm B}_4{\rm C}$ layer. A thermal break with this conductance is required in order to keep the heat conducted to the neutron shield reservoir low and encourage radiative cooling. The inner surface of the neutron shield was set to $600~^\circ {\rm C}$ to mimic the effect of the on-board shield reservoir. The results of this modeling are shown in Fig.~\ref{fig:neutron_heating} where the tungsten tiles in Reactor A reach a maximum steady state temperature of $1950$~K which is well below the design constraint of $2500$~K, but above the recrystallization temperature \cite{richou:2020, suslova:2014}. As long as the shield is maintained at these elevated temperatures it is possible that the onset of the degraded mechanical properties can be delayed until other forms of damage dominate. The energy that does not get radiated away at the surface is conducted inwards to the on-board reservoir. We calculate that $8$~\% of the neutron power deposited in the shield will need to be stored in the on-board reservoir and extracted during the docking procedure.

At these temperatures the lifetime of the tungsten tiles is also determined by the rate of thermal creep which acts to deform the tiles from their manufactured specifications. This rate is determined by the temperature, tungsten grain size, and material stress which can be controlled by varying the tungsten tile size. Up to a grain size of $100$~$\mu$m, stresses below $1$~MPa result in creep processes dominated by diffusion \cite{webb:2019} which can give part lifetimes in excess of a year. Determining the final tile size. and therefore their creep limited lifetime, requires a detailed design of the mounting mechanism of the tiles, which is outside the scope of this study.

At a preliminary level, the neutron shielding presented in this study satisfies all the requirements of an economically viable DT fusion power plant without the use of advanced materials. Further advancements in neutron shielding material science will only aid in the shield performance, allowing for more compact reactor designs and higher overall fusion power outputs.

\begin{figure*}[t!]
    \centering
    \includegraphics[width=1\linewidth]{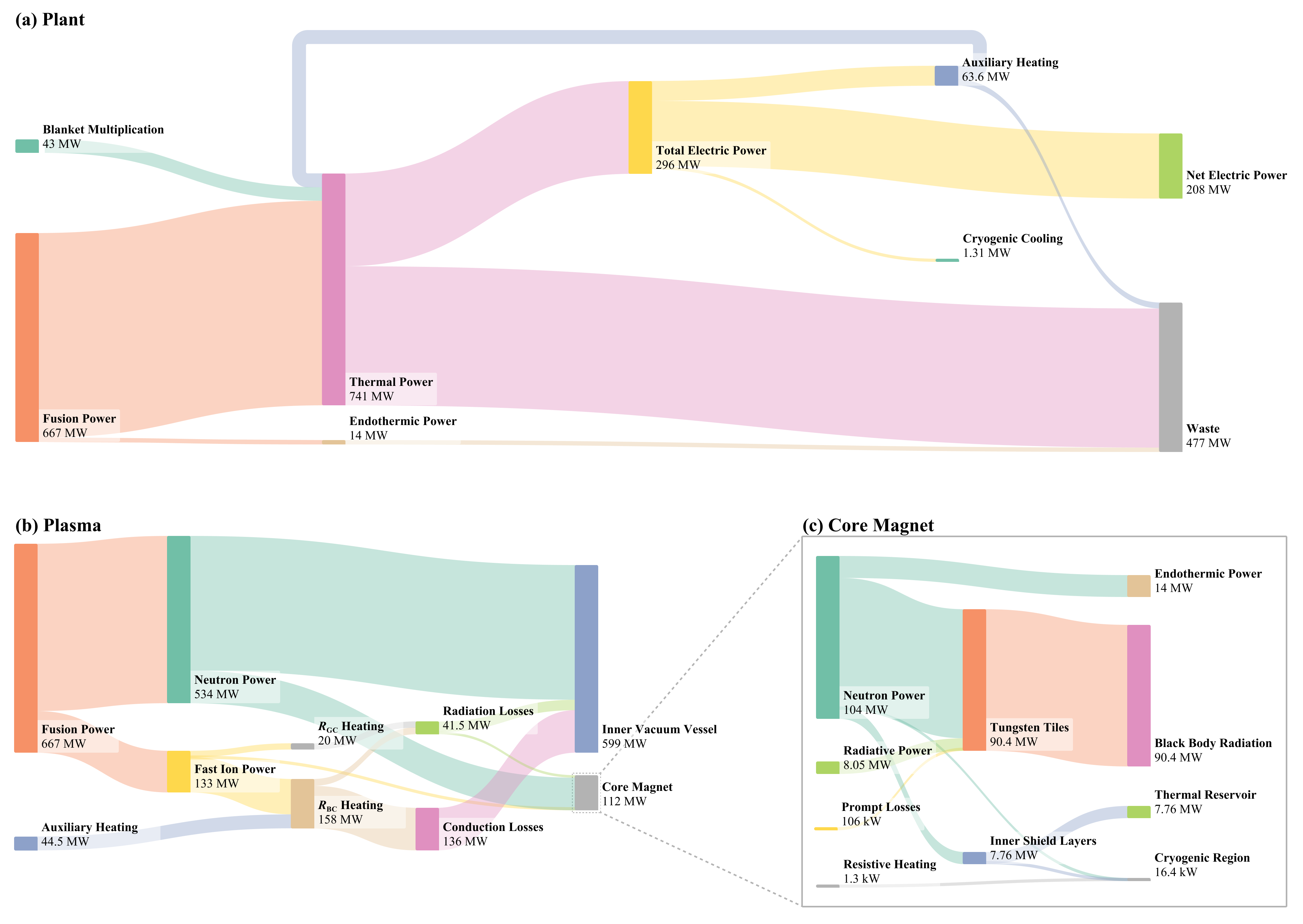}
    \caption{Sankey diagrams of the energy balance in Reactor A for (a) the full plant, (b) the plasma, and (c) the core magnet.}
    \label{fig:sankey}
\end{figure*}

\begin{table}[t!]
    \centering
    \resizebox{\columnwidth}{!}{
    \begin{tabular}{l c c l}
        \hline\hline
        Name & Reactor A & Reactor B & Units \rule{0pt}{4ex}\\ [1.5ex]

		\hline
		\textbf{Plasma} & & & \textbf{MW}\rule{0pt}{2.6ex} \\ [0.5ex]
		\hline
		$\mathfrak{R}_{\rm BC}$ $^{4}$He heating & $113$ & $40.2$ & \rule{0pt}{2.6ex} \\ [0.5ex]
		$\mathfrak{R}_{\rm GC}$ $^{4}$He heating & $20.0$ & $7.1$ &  \\ [0.5ex]
		Auxiliary heating & $44.5$ & $15.8$ &  \\ [0.5ex]
		Plasma Radiation & $-41.5$ & $-16.6$ &  \\ [0.5ex]
		Conduction & $-136$ & $-46.5$ &  \\ [0.5ex]
		\hline
		\textbf{First Wall} & \textbf{733} & \textbf{261} & \textbf{MW}\rule{0pt}{2.6ex} \\ [0.5ex]
		\hline
		Neutron Free Power & $430$ & $155$ & \rule{0pt}{2.6ex} \\ [0.5ex]
		Blanket Energy Multiplication & $43.0$ & $15.5$ &  \\ [0.5ex]
		Plasma Radiation & $33.4$ & $13.6$ &  \\ [0.5ex]
		Core Magnet Black Body Radiation & $90.4$ & $30.2$ &  \\ [0.5ex]
		Plasma Conduction & $136$ & $46.5$ &  \\ [0.5ex]
		\hline
		\textbf{Neutron Shield} & \textbf{7.76} & \textbf{2.65} & \textbf{MW}\rule{0pt}{2.6ex} \\ [0.5ex]
		\hline
		Neutron Free Power & $104$ & $34.7$ & \rule{0pt}{2.6ex} \\ [0.5ex]
		Endothermic Effects & $-14.0$ & $-4.9$ &  \\ [0.5ex]
		Plasma Radiation & $8.05$ & $2.99$ &  \\ [0.5ex]
		Prompt $^4$He Heating & $0.106$ & $0.058$ &  \\ [0.5ex]
		Black Body Radiation & $-90.4$ & $-30.2$ &  \\ [0.5ex]
		\hline
		\textbf{Core Magnet} & \textbf{16.4} & \textbf{9.97} & \textbf{kW}\rule{0pt}{2.6ex} \\ [0.5ex]
		\hline
		Neutron Heating & $13.5$ & $7.34$ & \rule{0pt}{2.6ex} \\ [0.5ex]
		Photon Heating & $0.59$ & $0.33$ &  \\ [0.5ex]
		Electrical Heating & $1.3$ & $1.3$ &  \\ [0.5ex]
		Conductive Heating & $1$ & $1$ &  \\ [0.5ex]
		\hline
		\textbf{Plant} & & & \textbf{MW}\rule{0pt}{2.6ex} \\ [0.5ex]
		\hline
		Fusion Power & $667$ & $237$ & \rule{0pt}{2.6ex} \\ [0.5ex]
		Thermal Power & $741$ & $264$ &  \\ [0.5ex]
		Total Electrical Power & $296$ & $106$ &  \\ [0.5ex]
		Cryogenic Cooling & $-1.31$ & $-0.798$ &  \\ [0.5ex]
		Plasma Heating Wall Power & $-63.6$ & $-22.6$ &  \\ [0.5ex]
		Net Electric Power & $208$ & $74.5$ &  \\ [0.5ex]

        \hline\hline
    \end{tabular}
    }
    \caption{Plant power fluxes to and from various systems/regions. The total power deposited in each region is given in bold.}
    \label{tab:EnergyBalance}
\end{table}

\subsection{Power Balance}\label{sec:a_power_balance}

\begin{figure*}[t!]
    \centering
    \includegraphics[width=1\linewidth]{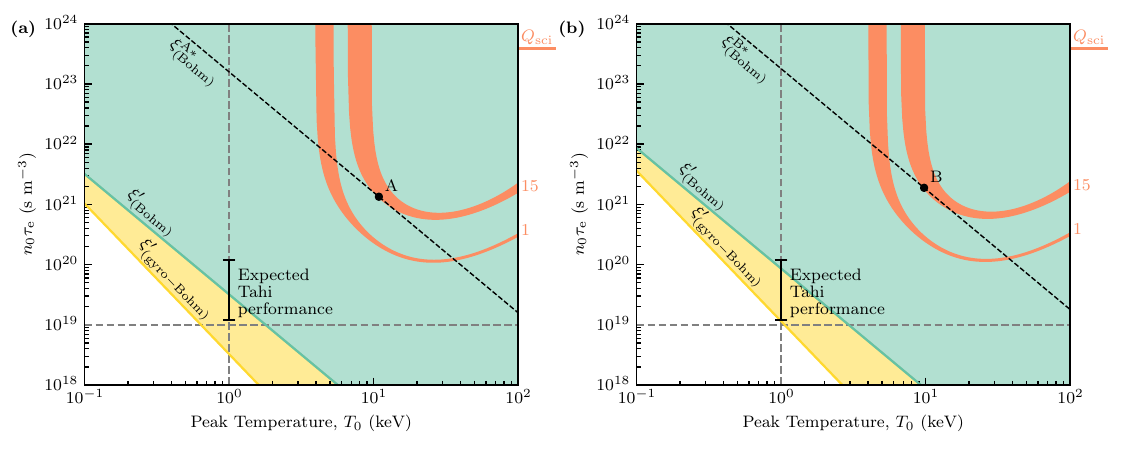}
    \caption{Performance acceptability plots for the optimized design points of (a) Reactor A ($208$~MWe) and (b) Reactor B ($75$~MWe). Both reactors are assumed to be $Q_{\rm sci}=15$ which then implies that a smaller device with a device index $\xi_{\alpha}'$ (Eq.~\eqref{eq:device_index}) must show double products greater than the back-scaled Bohm-like (green) or gyro-Bohm-like (yellow) operational contours in order for this assumption to be valid. The expected performance of the demonstration device, known as Tahi, is forward-calculated from LDX and used to illustrate the viability of each reactor. In both cases the expected performance band lies above the required gyro-Bohm performance. However, Reactor A being the larger device allows it to impose lower performance targets on Tahi, implying that a $Q_{\rm sci}=15$ Reactor A can be achieved with Bohm-like scaling.}
    \label{fig:tahi_scaling}
\end{figure*}

Combining the results from the previous sections the net electrical power of the reactor and the remaining full power balance can be calculated. The results of this calculation are provided in Table~\ref{tab:EnergyBalance} and visualized in Fig.~\ref{fig:sankey}. Starting in the plasma, the total heating is sourced from the auxiliary heating power and charged fusion products. The auxiliary power is assumed to be deposited entirely in the bad curvature region to enable some control over the location of the pressure peak. The heating from the fusion $\alpha$ particles is split between prompt losses to the surface of the core magnet neutron shield and the good and bad curvature regions. Of the $\alpha$ particles that are not lost to the neutron shield, it was assumed that $15$~\% of the $\alpha$ energy was deposited in the good curvature region with the remainder depositing their energy in the bad curvature region based on initial results from the ASCOT5 model in Section~\ref{sec:a_alpha_heating_results}. The energy deposited in the good curvature region is assumed to be converted to radiation and is included with the bremsstrahlung losses in the total radiation tally, $P_{\rm rad}$. All energy in the plasma is lost either through radiation or conduction to $\Psi_{\rm lcfs}$. The fraction of neutron power and radiated power that gets directly deposited in the first wall is determined by the OpenMC calculations discussed in Section~\ref{sec:a_neutron_shielding_results}. Extra energy is generated by neutron induced exothermic reactions in the Li$_2$O breeder blanket. The first wall also serves to capture the significant amounts of black body radiation emanating from the surface of the core magnet neutron shield, which indirectly captures the majority of the energy deposited in the shield. The bremsstrahlung and prompt losses to the neutron shield are assumed to be entirely balanced by black body radiation as they do not penetrate far into the shield volume. The deeper penetrating neutron power is then split between endothermic losses, black body radiation, and conduction to the thermal reservoir according to both the OpenMC model and the COMSOL neutron shield heat transfer model in Section~\ref{sec:a_neutron_shielding_results}. It is worth noting here that the endothermic effects in the neutron shield tungsten tiles removes $2~\%$ of the useful thermal power of the plant. This aids the problem of cooling the neutron shield, but reduces the thermal power of the overall plant. This is not only because of the direct energy loss but also because of the loss of exothermic reactions that would have been available if a tritium breeding blanket was mounted on the core magnet. However, the reduction in shield thickness granted by using tungsten over other materials more than makes up for this loss by reducing the overall size of the plant. 

The total thermal power is therefore the sum of the total power deposited in the first wall and the power deposited in the neutron shield reservoir. The total electrical power is then calculated assuming a $40~\%$ electrical conversion efficiency. The electrical power required to cool the magnet is calculated using an efficiency of $1.25$~\% (Section~\ref{sec:dp_cryogen}) which is applied to the total heating power in the cryogenic region. The total efficiency of the auxiliary heating systems is assumed to be $70$~\% which is then used to calculate the required total electrical power. Finally the net electrical power is calculated using Eq.~\eqref{eq:net_powaaahhh} to give a net output power of $208$~MW for Reactor A and $74.5$~MW for Reactor B.

\section{Discussion}\label{sec:a_scaling_results}

The final operating points are plotted in Fig.~\ref{fig:tahi_scaling} with their associated ``operating contours'' obtained by changing the core temperature at constant $\beta_0$ and device index, $\xi_{\alpha}$, as defined in Eq.~\eqref{eq:device_index}. Due to the temperature dependence of $\tau_{\rm e}$ and the imposed constraint on the plant overnight capital cost, these optimized reactors are designed such that their operational contours are tangent to, or close to tangent to, the $Q_{\rm sci}=15$ contour as this minimizes the scaling constant $k_{\rm \alpha}$ given the constraint on the overnight capital cost. Therefore, this optimization process has produced reactors with lower operating temperatures, in the range of $\sim10$~keV, than would be expected for a tokamak \cite{sorbom:2015, MANTA:2024} or stellarator \cite{lion:2025} DT fusion power plants. This is likely to change as more is understood about the scaling laws that govern the energy confinement time in a levitated dipole.

The assumption that these reactors will be $Q_{\rm sci}=15$ is only valid if a smaller demonstration device, which we will call Tahi, displays adequate plasma performance. Tahi will be of a similar physical size as OpenStar's current device, Junior \cite{chisholm:2026} (a $\sim 1$~m diameter core magnet inside a $\sim 5$~m vacuum vessel), with more than $1$~MW of available auxiliary heating power. The aim of Tahi is to achieve triple products in excess of $10^{19}$~keV\,s\,m$^{-3}$ with $1$~keV ions in the smallest possible form factor. The detailed design and final operating point of Tahi will be covered in a future publication. 

In order for these reactors to be $Q_{\rm sci}=15$, Tahi will need to show double products above the required Bohm-like and gyro-Bohm-like scaled operational contours shown in Fig.~\ref{fig:tahi_scaling} calculated using Eq.~\eqref{eq:device_index}. For core temperatures of $1$~keV, Bohm-like scaling would require Tahi to show double products in excess of $3.23\times10^{19}$~s\,m$^{-3}$ for Reactor A and $8.69\times10^{19}$~s\,m$^{-3}$ for Reactor B. In both cases if Tahi were to show gyro-Bohm-like scaling, the requirement on the double product would be less than $10^{19}$~s\,m$^{-3}$ therefore implying the reactors shown here would be $Q_{\rm sci}>15$. 

The confinement time scaling in Eq.~\eqref{eq:device_index} can also be applied to data from the LDX device \cite{davis:2014}, which had a measured confinement time of $\tau_{\rm e}=14.5$~ms, to obtain a bound on the expected performance of Tahi. Conservatively applying a Bohm-like scaling from this device implies that a $10^{19}$~keV\,s\,m$^{-3}$ triple product can be achieved with a $\sim11$~T core magnet. Additionally, LDX confined a plasma that was susceptible to significant charge exchange losses and as such was never ionized all the way to $\Psi_{\rm lcfs}$ \cite{boxer:2010}. This, along with other possible effects caused by the drastic increase in available heating power over previous experiments, will only aid the plasma performance in Tahi further implying that Bohm-like scaling from LDX should provide a conservative lower bound. Therefore, we assume the upper performance bound to be 10 times the calculated lower bound as shown in Fig.~\ref{fig:tahi_scaling}. As a point of comparison, a similar analysis on historical tokamak data yields an uplift factor of between 40 and 100 \cite{wurzel:2022}. It then becomes clear that some increase in performance over purely Bohm-like scaling is needed in order for either reactor to be $Q_{\rm sci}\geq15$ given Bohm-like scaling from LDX, with Reactor B needing a more significant increase. However, the modest performance gain required for Reactor A would likely be achieved by fully ionizing the plasma in Tahi.

In the event that Tahi does not reach the double products outlined above, the reactors presented here would be $Q_{\rm sci}<15$. This would in turn either require an increase in the LCOE or the size, and therefore overnight capital cost, of the plant beyond the values set as constraints in this optimization. However, it is more likely, given the conservative scalings we have used, that Tahi will show the performance required for a $Q_{\rm sci}=15$ Reactor A in the Bohm-like scaling case. Additionally, if Tahi were to show gyro-Bohm-like scaling or outperform the Bohm-like targets then this would allow for smaller, more capitally efficient plants such as Reactor B to be built.

\section{Conclusions}\label{sec:conclusions}

This study has proposed two design points for first of a kind levitated dipole DT fusion reactors, a larger Reactor A with a total thermal power of $667$~MW ($208$~MWe) and another smaller Reactor B producing $237$~MW of thermal power ($75$~MWe). The levitated dipole allows for simple magnet geometries and stable steady state operation without damaging disruption-like events. The lack of a fusion relevant dipole experiment and the engineering design challenges associated with levitating a superconducting magnet within a fusing DT plasma have, until this study, been seen as the main detractor to the concept. 

This work presents high level designs for two critical components of the levitated dipole that were previously thought to be impractically difficult: The large, high field core magnet and its neutron shield. The core magnet leverages advancements in superconducting technology to both produce strong magnetic fields ($>20$~T) and maintain a steady state operating current through the use of an on-board superconducting power supply. The shape of the core magnet coil was optimized to reduce mechanical strains to $<0.4$~\% with the help of a structural over-band. An internal low-field region was also created to house the superconducting power supply and other required on-board systems. These factors allow the presented core magnet structure to be built using traditional methods and materials. The neutron shielding we present here successfully attenuates the DT neutron flux down to acceptable levels, only depositing $\sim 10$~kW of heating into the cryogenic region and allowing for partial magnet lifetimes of $1$~year and full magnet lifetimes on the order of $10$~years. The neutron shield itself is constructed from layers of tungsten and ${\rm B}_4{\rm C}$, which allows the neutron shield to operate at the extreme temperatures ($1950$~K) required for radiative cooling to the first wall. 

The large vacuum vessel will be constructed using a two layer approach. The outer layer will support close to the full loads of the vacuum and core magnet and will be built from reinforced concrete, enabling the large vessel diameters required for a levitated dipole fusion rector. This construction also allows excess room for a tritium breeding blanket mounted to the outside of the inner vacuum vessel. With assistance from neutron reflective core magnet shield, the space available for the tritium breeder blanket allows for tritium breeding ratios in excess of $1.1$ without the use of expensive molten salts and neutron multipliers.  

These designs show that the engineering of a practically sized levitated dipole reactor is feasible. The presented reactors set performance requirements that a sub-scale demonstration device will need to exceed in order to fully validate the designs. This analysis in turn showed that levitated dipoles will likely need to show better than Bohm-like scaling in order for the reactor designs presented in this study to be valid. Both reactors were also designed with economic constraints in mind, leveraging the inherent modularity of the dipole which allows for easy replacement of the core magnet to compensate for the shorter coil lifetime. Hence, these design points are not only physically feasible, but they are also expected to be economically attractive.

\appendix

\section{Geometry Factors}\label{app:geometry_factors}

In Section~\ref{sec:m_cost_function} we have derived the 0D power balance and confinement time relations using plasma averaged quantities to capture the effect of the highly peaked pressure profile. However, it is more convenient to index these relations using values measured at the pressure peak. To convert between peak and average values, we utilize the formalism of Wurzel and Hsu \cite{wurzel:2022} to define factors that capture the geometric effect of the plasma:
\begin{equation}
    \lambda \equiv \frac{\langle A \rangle}{A_0}.
\end{equation}
where $\langle A\rangle$ is the volume average of quantity $A$ and $A_0$ is the value of $A$ measured at the pressure peak. The quantities of interest for a power balance are the fusion power density
\begin{equation}
    \lambda_{\rm f} \equiv \frac{\langle n^2 \langle \sigma v\rangle_v\rangle}{n_0^2 \langle \sigma v\rangle_{v,0}},
\end{equation}
bremsstrahlung power density
\begin{equation}
    \lambda_{\rm b} \equiv \frac{\langle n^2 \sqrt{T}\rangle}{n_0^2 \sqrt{T_0}},
\end{equation}
and the conductive power losses
\begin{equation}
    \lambda_{\rm \kappa} \equiv \frac{\langle nT\rangle}{n_0T_0}.
\end{equation}
\begin{figure}[!b]
    \centering
    \includegraphics[width=1\linewidth]{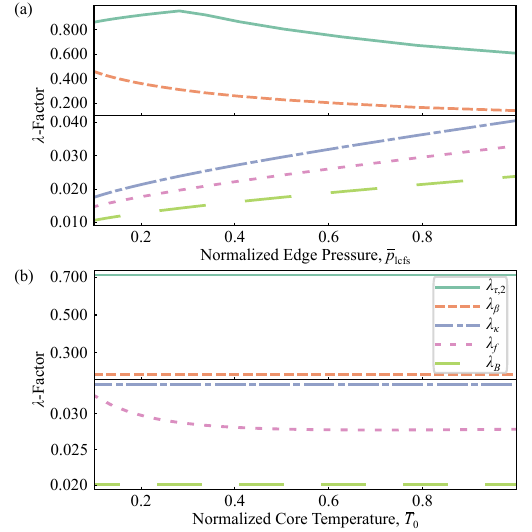}
    \caption{Dependence of the value of the geometry factors discussed in the text on (a) the plasma pressure at the last closed flux surface and (b) the core temperature of the plasma. The model used here is the same as in Fig.~\ref{fig:eqbs}.}
    \label{fig:lambda_factors}
\end{figure}
Also useful for the study of dipoles are the geometric factors for local plasma $\beta$:
\begin{equation}
    \lambda_{\rm \beta} \equiv \frac{\langle \beta\rangle}{\beta_0},
\end{equation}
and for the energy confinement time:
\begin{equation}
    \lambda_{\rm \tau,\alpha} \equiv \frac{\langle \Omega_{\rm i}\rho_{*}^\alpha\rangle_{\rm mid}}{\Omega_{\rm i, 0}\rho_{*, 0}^\alpha},
\end{equation}
where, as discussed in Section~\ref{sec:m_cost_function}, in this last case the average is taken only over the outer midplane as the transport there dominates:
\begin{equation}
    \langle \Omega_{\rm i}\rho_{*}^\alpha\rangle_{\rm mid} \equiv \frac{1}{a}\int_{R_0}^{R_{\rm lcfs}}\Omega_{\rm i}\rho_*^\alpha~\d r.
\end{equation}
Typically these values, aside from $\lambda_{\rm f}$ due to the strong dependence on $T$ in $ \langle \sigma v\rangle_v$, are assumed to be constant for all operating points of any given device. While this remains true for variation in temperature in a dipole, Fig.~\ref{fig:lambda_factors} shows that changing the pressure of the device, and therefore also the device $\beta$, also has an effect on the geometry factors. For the standard factors $\lambda_{\rm f}$, $\lambda_{\rm b}$, and $\lambda_{\rm \kappa}$ the dependence captures the overall expansion of the plasma as discussed in Section~\ref{sec:dp_large_beta}. The $\lambda_{\rm \beta}$ and $\lambda_{\rm \tau}$ factors on the other hand show very different behavior. Increased edge pressures result in more peaked local $\beta$ profiles, as expected from the lack of a limit on $\beta_0$ as the plasma expands. The large value of $\lambda_{\rm \tau}$ in the case of Bohm-like scaling ($\lambda_{\rm \tau, 2}$) indicates that the quantity $\Omega_{\rm i}\rho_{*}^2$ is close to uniform along the plasma outer midplane. 

Using these geometry factors, Eqs.~\eqref{eq:device_index} and \eqref{eq:taue_from_q} can be equivalently expressed as:
\begin{equation}
    \tau_{\rm e}=\frac{k_{\rm \alpha}}{\lambda_{\rm \tau,\alpha}}\frac{a^{\alpha}B_0^{\alpha-1}}{T_0^{\alpha/2}},
\end{equation}
and
\begin{equation}
    \tau_{\rm e} = \frac{\frac{3}{2}(1+\bar{Z})\lambda_{\rm \kappa}\frac{T_0}{n_0}}{\frac{E_{\rm f}}{4}\left(f_{\rm bcr}f_{\alpha}+\frac{\eta_{\rm h}}{Q_{\rm sci}}\right)\lambda_{\rm f}\langle\sigma v\rangle_{v} - C_{\rm B}Z_{\rm eff}\bar{Z}^2\lambda_{\rm b}\sqrt{T_0}}.
\end{equation}
%


\bibliography{references}
\end{document}